\documentclass[aps,prb,onecolumn,longbibliography,showpacs,10pt]{revtex4-1}
\usepackage{graphicx,bm,wasysym,color}

\def\be{\begin{equation}}  
\def\ee{\end{equation}}  
\def\ba{\begin{eqnarray}}  
\def\ea{\end{eqnarray}}  
\def\bc{\begin{center}}  
\def\ec{\end{center}}  
\def\p{\partial}  
\def\sgn{{\rm sgn}}

\begin{document}

\title{Nonperturbative quasi-classical theory of the nonlinear electrodynamic response of graphene}

\author{S. A. Mikhailov}
\email[Email: ]{sergey.mikhailov@physik.uni-augsburg.de}
\affiliation{Institute of Physics, University of Augsburg, D-86135 Augsburg, Germany}

\date{\today}

\begin{abstract}
An electromagnetic response of a single graphene layer to a uniform, arbitrarily strong electric field ${\bm E}(t)$ is calculated by solving the kinetic Boltzmann equation within the relaxation-time approximation. The theory is valid at low (microwave, terahertz, infrared) frequencies satisfying the condition $\hbar\omega\lesssim 2E_F$, where $E_F$ is the Fermi energy. We investigate the saturable absorption and higher harmonics generation effects, as well as the transmission, reflection and absorption of radiation incident on the graphene layer, as a function of the frequency and power of the incident radiation and of the ratio of the radiative to scattering damping rates. We show that the optical bistability effect, predicted in Phys. Rev. B \textbf{90}, 125425 (2014) on the basis of a perturbative approach, disappears when the problem is solved exactly. We show that, under the action of a high-power radiation ($\gtrsim 100$ kW/cm$^2$) both the reflection and absorption coefficients strongly decrease and the layer becomes transparent.
\end{abstract}


\maketitle


\section{Introduction\label{sec:intro}}

One of the most distinctive features of graphene, important for its electronic and optoelectronic applications, is the linear energy dispersion ${\cal E}_\pm ({\bm p})=\pm v_F|{\bm p}|$ of quasi-particles -- electrons and holes -- in this material \cite{Neto09}. It was shown \cite{Mikhailov07e}, within the quasi-classical approach, that the linear energy dispersion of graphene electrons should lead to its strongly nonlinear electrodynamic response. Indeed, under the action of a time-dependent electric field proportional to $\cos \omega t$ the electron momentum should oscillate, according to the Newton's equation of motion, as $\sin\omega t$. In conventional materials with the parabolic energy dispersion the velocity, and hence the current, are proportional to the momentum, therefore the current oscillates with the same frequency $\omega$. In contrast, in graphene the velocity ${\bm v}={\bm\nabla}_{\bm p}{\cal E}_\pm ({\bm p})=v_F{\bm p}/|{\bm p}|$ is proportional not to the momentum but, roughly, to the sign of the momentum. As a result the time dependence of the current $j$ has a step-like form,
\be 
 j\propto en_s v_F\sgn (\sin\omega t)=en_s v_F\frac 4\pi\left( \sin\omega t)+\frac 13\sin(3\omega t)+\dots\right),
\ee
and contains higher frequency harmonics \cite{Mikhailov07e} (here $n_s$ is the surface electron density in graphene).

Graphene is thus intrinsically a strongly nonlinear material, and all the variety of nonlinear phenomena -- harmonics generation, frequency mixing, saturable absorption, and so on, -- should be observed in graphene in relatively low electric fields. The prediction \cite{Mikhailov07e} was experimentally confirmed in many papers where harmonics generation \cite{Dragoman10,Bykov12,Kumar13,Hong13,An13,An14,Lin14}, four-wave mixing  \cite{Hendry10,Hotopan11,Gu12}, saturable absorption and Kerr effect \cite{Zhang12,Popa10,Popa11,Vermeulen16}, plasmon related nonlinear phenomena \cite{Constant16}, and other effects \cite{Mics15} were observed. This also stimulated further theoretical studies of different nonlinear electrodynamic effects in graphene \cite{Mikhailov08a,Mikhailov09a,Mikhailov09b,Dean09,Dean10,Ishikawa10,Mikhailov11c,Jafari12,Mikhailov12c,Avetissian13,Mikhailov13c,Cheng14a,Cheng14b,Yao14,Smirnova14,Peres14,CoxAbajo14,CoxAbajo15,Savostianova15,Cheng15,*Cheng16,Mikhailov16a,Mikhailov16b,Rostami16a,CoxAbajo16,MariniAbajo16,Sharif16,Sharif16a,Cheng16b,Wang16,Tokman16,Rostami16b}, for recent reviews see \cite{Glazov14,Hartmann14}. 
Theoretically, analytical results for the nonlinear electromagnetic response of a uniform graphene layer were obtained both at low (microwave, terahertz, $\hbar\omega\lesssim 2E_F$, Refs. \cite{Mikhailov07e,Mikhailov08a,Mikhailov11c,Yao14,Peres14}), and high (infrared, optical, $\hbar\omega\gtrsim 2E_F$, Refs. \cite{Ishikawa10,Cheng14a,Cheng14b,Cheng15,*Cheng16,Mikhailov16a,Rostami16a,MariniAbajo16,Wang16,Cheng16b,Tokman16}), frequencies; here $E_F$ is the Fermi energy of graphene electrons (or holes) and $\omega$ is the typical radiation frequency. In the latter case the inter-band electronic transitions have to be taken into account and the problem requires a quantum-mechanical treatment. In the former case it is sufficient to consider only the intra-band transitions and the problem can be solved within the quasi-classical Boltzmann equation.

In most so far published theoretical papers the nonlinear electromagnetic response of graphene was studied within the perturbation theory. In this paper we study the low-frequency ($\hbar\omega\lesssim 2E_F$) electromagnetic response of graphene \textit{non-perturbatively}. We solve the quasiclassical Boltzmann equation  in the relaxation time ($\tau$) approximation,
\be 
\frac{\p f(\bm p,t)}{\p t}+F_x(t)\frac{\p f(\bm p,t)}{\p p_x}=-\frac{f(\bm p,t)-f_0(\bm p)}{\tau}\label{be},
\ee
$F_x(t)=-eE_x(t)$, not assuming that the ac electric field $E_x(t)$ acting on graphene electrons is weak; here $f_0(\bm p)$ is the equilibrium (Fermi-Dirac) distribution function. Having found the non-perturbative electron distribution function we then calculate the nonlinear current and analyze the saturable absorption, harmonic generation and some other nonlinear effects.

The $\tau$-approximation (\ref{be}) that we use in this work offers a simple but efficient way to take into account the charge carrier scattering processes. It often allows one to get even \textit{analytical} solutions of complicated  nonlinear response problems (see Refs. \cite{Cheng14b,Yao14,Peres14,Cheng15,Mikhailov16a,Mikhailov16b,Cheng16b}, as well as this work). The time  $\tau$ in (\ref{be}) is a phenomenological parameter which may depend on temperature, chemical potential and radiation power. The relaxation-time approximation may fail near resonances related to inter-band transitions, in particular, when a nonlinear response of intrinsic graphene ($E_F=0$) is considered \cite{Romanets10,Romanets11}. In the present work we focus on the quasi-classical frequency range $\hbar\omega\lesssim 2E_F$, where the inter-band transitions are not the case, which justifies the use of the approximation (\ref{be}). 

Among the nonlinear phenomena which turned out to be useful to study within the nonperturbative theory is the so called \textit{optical bistability}. This effect was predicted in graphene at terahertz frequencies ($\hbar\omega\lesssim 2E_F$) in a recent work \cite{Peres14}. In that paper the authors considered the incidence of radiation (with the frequency $\omega$ and the intensity ${\cal J}_{i}$) on a graphene layer, and calculated the intensity of the transmitted wave at the same frequency ${\cal J}_{t}$. Solving the Boltzmann equation (\ref{be}) they showed that, at certain (large) values of the parameter
\be 
\beta =\left(\frac{2\alpha E_F}{\hbar\omega}\right)^2,
\ee
the dependence ${\cal J}_{t}({\cal J}_{i})$ has an $S$-shaped form, i.e., the function ${\cal J}_{t}({\cal J}_{i})$ is multivalued; here $\alpha=e^2/\hbar c$ is the fine structure constant. This result was derived in the collisionless approximation, $\omega\tau\gg 1$, and in \textit{low} electric fields, meaning that the solution of the Boltzmann equation was expanded in powers of the parameter
\be 
{\cal F}_\omega=\frac{eE_0}{\hbar\omega k_F},\label{fieldpar}
\ee 
taking the third ($\propto {\cal F}_\omega^3$) and fifth ($\propto {\cal F}_\omega^5$) order terms into account; here $E_0$ is the electric field amplitude and $k_F$ is the Fermi wave-vector. 

The results of Ref. \cite{Peres14} cause a few questions.  First, the low-frequency response of graphene was investigated, in the collisionless approximation $\omega\tau\gg 1$, in Ref. \cite{Mikhailov07e}. In the paper \cite{Mikhailov07e} the right-hand side of Eq. (\ref{be}) was assumed to be zero from the outset, while in Ref. \cite{Peres14} equation (\ref{be}) was first solved with a finite right-hand side and \textit{after that} the limit $\omega\tau\gg 1$ was taken. The results of Refs. \cite{Mikhailov07e} and \cite{Peres14} coincide for the third-order response (for the $3\omega$-Fourier component of the current), but differ (by a factor of three) for the $\omega$-Fourier component. The question why the two different methods give different results \textit{under the same condition} $\omega\tau\gg 1$ remained unclear (a short comment on page 3 of Ref. \cite{Peres14} does not actually explain this contradiction).

Second, the $S$-shaped ${\cal J}_{t}({\cal J}_{i})$--characteristics was obtained in Ref. \cite{Peres14} by using the perturbative theory in the electric field parameter ${\cal F}_\omega$, which supposes, strictly speaking, that ${\cal F}_\omega$ should be small, ${\cal F}_\omega\ll 1$. But the characteristic point of the $S$-shaped ${\cal J}_{t}({\cal J}_{i})$--curve, in which the derivative $d{\cal J}_{t}/d{\cal J}_{i}$ becomes infinite, $d{\cal J}_{t}/d{\cal J}_{i}=\infty$, lie at ${\cal F}_\omega\simeq 1$, see Figure 3 in Ref. \cite{Peres14}. This causes some doubts in the validity of the prediction; the question, whether the predicted bistability survives if to solve the problem exactly, not expanding the result in powers of ${\cal F}_\omega$, remained unanswered.

The third question concerns the applicability area of the predicted effect. As seen from Figure 3 of Ref. \cite{Peres14}, the optical bistability takes place only if $\beta\gtrsim 2$. This means that the condition $\hbar\omega\lesssim  \alpha E_F$ should be satisfied. On the other hand, $\omega\tau$ should be much larger than unity. This leaves a narrow window for the frequency, $\alpha E_F/\hbar\gtrsim\omega\gg 1/\tau$ and imposes rather strong restrictions on the Fermi energy, $E_F\gg 137\hbar/\tau$, and the mean free path $l=v_F\tau$, $k_F l\gg 137$. The question arises, whether and how the predicted effect is modified if to do not assume that $\omega\tau\gg 1$ and $k_F l\gg 137$.

In this paper we answer the three above formulated questions. We solve the problem exactly, \textit{not assuming} that the frequency parameter $\omega\tau$ is large and the field parameter ${\cal F}_\omega$ is small. We show that, if to solve the problem exactly (non-perturbatively), the ``optical bistability'' effect disappears. The ${\cal J}_{t}({\cal J}_{i})$--characteristics of graphene remains a nonlinear but single-valued function which physically corresponds to the saturable absorption but \textit{not} to the optical bistability. We also show that at small values of the frequency parameter $\omega\tau$ (the limit that was not considered in Ref. \cite{Peres14}) the nonlinear features are in fact more pronounced than in the limit $\omega\tau\gg 1$.

The paper is organized as follows. In Section \ref{sec:theory} we formulate the non-perturbative nonlinear response problem and solve it. In Section \ref{sec:results} we analyze the obtained results in several different cases including the Kerr and harmonics generation effects. In Section \ref{sec:summary} the results are summarized and conclusions are drawn. Some technical details are given in the Appendixes.  

\section{Theory\label{sec:theory}}

\subsection{Formulation of the problem}

We consider a homogeneous two-dimensional (2D) electron gas under the action of a uniform electric field $E_x(t)$. The 2D layer occupies the plane $z=0$ and the spectrum of electrons in it, ${\cal E}({\bm p})\equiv {\cal E}(p_x,p_y)$, can be parabolic, like in a semiconducting GaAs quantum well, or linear, like in semimetallic graphene. Further, we consider an experimentally relevant situation, when the electric field $E_x(t)$ is zero at $t<0$ and is switched on at $t=0$ being, in general, an arbitrary function $\tilde E_x(t)$ at $t>0$, $E_x(t)=\theta(t) \tilde E_x(t)$. The distribution of electrons in the momentum space is described, respectively, by the Fermi-Dirac function
\be 
f_0({\bm p})\equiv f_0(p_x,p_y)=\left[1+\exp\left(\frac {{\cal E}({\bm p})-\mu}T\right)\right]^{-1},
\ee
at $t<0$ and by Boltzmann equation (\ref{be}) at $t>0$; here $\mu$ is the chemical potential and $T$ is the temperature. The scattering of electrons is taken into account within the momentum relaxation time approximation, Eq. (\ref{be}), where $\tau$ is assumed to be energy independent. Our task is to find the distribution function $f(\bm p,t)\equiv f(p_x,p_y,t)$, which satisfies Boltzmann equation (\ref{be}) at $t>0$ and the initial condition 
\be 
f(p_x,p_y,t=0)=f_0(p_x,p_y)\label{initf}
\ee
at $t=0$, not imposing any restriction of the value of the scattering parameter $\omega\tau$ and not expanding the solution in powers of the electric field.

\subsection{Solution of Boltzmann equation}

The formulated problem is often solved by the method of characteristics, see, e.g., Ref. \cite{Peres14}. We use the Fourier technique which, in our opinion, is more transparent. First, we notice that the momentum component $p_y$ in Eq. (\ref{be}) is a parameter, and the functions $f_0(p_x,p_y)$ and $f(p_x,p_y,t)$ tend to zero at $p_x\to \pm \infty$. Therefore we can expand these functions in Fourier integrals over $p_x$:
\be 
f(p_x,p_y,t)=\int_{-\infty}^\infty ds e^{isp_x}\tilde f(s,p_y,t),
\label{Fourf}
\ee
\be 
f_0(p_x,p_y)=\int_{-\infty}^\infty ds e^{isp_x}\tilde f_0(s,p_y).
\label{Fourf0}
\ee
Substituting these expansions in Boltzmann equation (\ref{be}) we get an inhomogeneous ordinary differential equation for the function $\tilde f(s,p_y,t)$ 
\be 
\frac{d \tilde f(s,p_y,t)}{d t}+\left(is F(t)+\frac{1}{\tau}\right)\tilde f(s,p_y,t)=\frac{\tilde f_0(s,p_y)}{\tau},\ \ \ t>0, \label{inhomo}
\ee
with the initial condition 
\be 
\tilde f(s,p_y,t=0)=\tilde f_0(s,p_y). \label{init}
\ee
($s$ and $p_y$ are parameters here). Eq. (\ref{inhomo}) can be solved by the separation of variables and the variation of constants techniques, which leads to the general solution 
\be 
\tilde f(s,p_y,t)=A_0
e^{-\gamma t-is \int_{-\infty}^t F(t') dt'}
+
\gamma \tilde f_0(s,p_y) 
\int_{-\infty}^te^{-\gamma (t-t')-is \int_{t'}^t F(t'') dt''}d t' ,
\label{res2}
\ee
where $\gamma=1/\tau$ and $A_0$ is the integration constant. At $t<0$ the force $F_x(t)$ equals zero, hence, the integrals $\int_{-\infty}^t F(t') dt'$ and $\int_{t'}^t F(t'') dt''$ vanish and we get
\be 
\tilde f(s,p_y,t<0)= A_0 e^{-\gamma t}+
\tilde f_0(s,p_y). \label{res1}
\ee
Comparing (\ref{res1}) with (\ref{init}) we see that $A_0=0$. Substituting now  (\ref{res2}) into (\ref{Fourf}) we obtain, see Appendix \ref{Deriv}:
\be 
f(p_x,p_y,t)=
\gamma
\int_{-\infty}^te^{-\gamma (t-t')}
 f_0\left(p_x - \int_{t'}^t F(t'') dt'',p_y\right)d t'.
\label{resultBE}
\ee
The function (\ref{resultBE}) satisfies Boltzmann equation (\ref{be}) and the initial condition (\ref{initf}). At $t>0$ it can be rewritten as \cite{Ignatov76} (Appendix \ref{Deriv}):
\be 
f(p_x,p_y,t)=
e^{-\gamma t}
 f_0\left(p_x - \int_{0}^t \tilde F(t'') dt'',p_y\right)
+\gamma
\int_{0}^te^{-\gamma (t-t')}
 f_0\left(p_x - \int_{t'}^t \tilde F(t'') dt'',p_y\right)d t' .
\label{resultSplitted}
\ee


The difference between the results of Refs. \cite{Mikhailov07e} and \cite{Peres14} can now be clarified. In Ref. \cite{Mikhailov07e} the limit $\tau\to\infty$ was taken from the outset. Assuming $\gamma=1/\tau\to 0$ in (\ref{resultSplitted}) we get
\be 
\lim_{\gamma\to0}f(p_x,p_y,t)=
 f_0\left(p_x - \int_{0}^t \tilde F(t'') dt'',p_y\right).
\label{resultMikhailov}
\ee
This result coincides with the one obtained in Ref. \cite{Mikhailov07e} (if to assume that the force $\tilde F(t)$ contains only one frequency harmonic). One sees that the limit $\gamma\to0$ ($\tau\to\infty$) actually means $\gamma t\ll 1$ or $t\ll \tau$, i.e., the result of Ref. \cite{Mikhailov07e}, Eq. (\ref{resultMikhailov}), is valid under the conditions
\be 
\omega\tau\gg 1\textrm{ and }t\ll \tau\ \ \textrm{ in Ref. \cite{Mikhailov07e}}.
\label{condMikh}
\ee

In the opposite limit, $\gamma t\gg 1$, Eq. (\ref{resultSplitted}) gives 
\be 
\lim_{\gamma t\to\infty} f(p_x,p_y,t)=
 \int^{\infty}_0 e^{-\xi} f_0\left(p_x- \int_{t-\xi\tau}^t \tilde F(t'') dt'',p_y\right)   d \xi,
\label{resultPeres}
\ee
see Appendix \ref{Deriv}. This result coincides with the one obtained in Ref. \cite{Peres14} (if to assume that the force $\tilde F(t)$ contains only one frequency harmonic). It is thus valid at 
\be 
\omega\tau\gg 1\textrm{ and }t\gg \tau\ \ \textrm{ in Ref. \cite{Peres14}}.
\label{condPeres}
\ee
Thus, being both obtained under the condition $\omega\tau\gg 1$, the results of Refs. \cite{Mikhailov07e} and \cite{Peres14} are valid in different time intervals, Eqs. (\ref{condMikh}) and (\ref{condPeres}), respectively. Below, we will only consider the steady-state solution (\ref{resultPeres}), which is valid at $t\gg\tau$ and which we will simply call $f(p_x,p_y,t)$. 

\subsection{Response to a monochromatic excitation}

If $\tilde F(t)$ is a periodic function with a period $T$, the distribution function $f(p_x,p_y,t)$, Eq. (\ref{resultPeres}), is also periodic with the same period. Assume now that the external force is given by a monochromatic sine function,
\be 
\tilde F(t)=-eE_0\sin\omega t,\label{ext-field}
\ee
and calculate the induced electric current 
\be 
j_x(t)=-\frac{e}{S}\sum_{\bm p\sigma v} \frac{\p {\cal E}(\bm p)}{\p p_x} f(p_x,p_y,t)
=
-\frac{e}{S}\sum_{\bm p\sigma v} \frac{\p {\cal E}(\bm p)}{\p p_x} \int^{\infty}_0 e^{-\xi} f_0\Big(p_x- p_0(t,\xi\tau),p_y\Big)   d \xi.\label{current}
\ee
Here $S$ is the sample area, $\sigma$ and $v$ are the spin and valley (if applicable) quantum numbers, and 
\be 
p_0(t,\xi\tau)\equiv \int_{t-\xi\tau}^t \tilde F(t') dt'
=\frac{eE_0}\omega \left[\cos \omega t -\cos\omega (t-\xi\tau)\right]
.\ee
From (\ref{current}) one immediately sees that, if the spectrum of electrons is parabolic, the system response is linear at arbitrary values of the electric field. Substituting $\p {\cal E}/\p p_x=p_x/m$ and replacing the variable $p_x-p_0(t,\xi\tau)\to p_x$, we get
\be 
j_x(t)=
-\frac{en_s}{m}\int^{\infty}_0 e^{-\xi}  d \xi p_0(t,\xi\tau)=  \frac{n_se^2\tau}{m}E_0\frac {\sin\omega t -\omega\tau \cos\omega t }{1+(\omega\tau)^2}, \label{DrudeParab}
\ee
where
\be 
n_s=\frac{1}{S}\sum_{\bm p\sigma v}  f_0(\bm p)  
\ee
is the two-dimensional (2D) equilibrium electron density. Introducing the complex-valued linear-response Drude conductivity 
\be 
\sigma_D(\omega)=\frac{\sigma_0}{1-i\omega\tau},\ \ \ \sigma_0=\frac{n_se^2\tau}{m},\label{DrudeConduct}
\ee
we see that Eq. (\ref{DrudeParab}) can be rewritten as
\be 
j_x(t)= \Big(\sigma'_D(\omega)\sin\omega t -\sigma''_D(\omega) \cos\omega t \Big)E_0, \label{DrudeParabA}
\ee
where $'$ and $''$ mean the real and imaginary parts of $\sigma_D(\omega)$. 
This is the standard linear Drude response of a 2D gas of massive electrons. 

Now consider graphene in which electrons have the linear energy dispersion 
\be 
{\cal E}(\bm p)=v_F|\bm p|=v_F\sqrt{p_x^2+p_y^2},
\label{spectrum}
\ee
where $v_F\approx 10^8$ cm/s is the  Fermi velocity. Substituting (\ref{spectrum}) into (\ref{current}) we get 
\be 
j_x(t)=
-\frac{eg_sg_vv_F}{(2\pi\hbar)^2}\int^{\infty}_0 e^{-\xi}  d \xi 
\int\int  \frac{p_xdp_xdp_y}{\sqrt{p_x^2+p_y^2}} f_0\left(p_x- p_0(t,\xi\tau),p_y\right),  
\label{curr1}
\ee
where $g_s=g_v=2$ are the spin and valley degeneracies. Further, assuming that the temperature is low, $T/ E_F\to 0$, we transform Eq. (\ref{curr1}) to the form 
\be  j_x(t)= -\frac{eg_sg_vv_Fp_F^2}{(2\pi\hbar)^2}\int^{\infty}_0  {\cal I}\left[\Pi({\cal F}_\omega,\omega t,\omega\xi\tau)\right]e^{-\xi}  d \xi,  \label{curr1a} \ee
where $p_F=\hbar k_F$ is the Fermi momentum,
\be 
{\cal I}(\Pi)=
2 \int_{0}^{1}dy \left(\sqrt{1+\Pi^2 +2\Pi\sqrt{1-y^2}} -\sqrt{1+\Pi^2 -2\Pi\sqrt{1-y^2}} \right),\label{integralI-peresform}
\ee 
and the function
\be 
\Pi\equiv \Pi({\cal F}_\omega,\omega t,\omega\xi\tau)=\frac{p_0(t,\xi\tau)}{p_F}={\cal F}_\omega\left[\cos \omega t -\cos\omega (t-\xi\tau)\right]\label{funcK}
\ee
is  proportional to ${\cal F}_\omega$, see Eq. (\ref{fieldpar}). The electric field parameter ${\cal F}_\omega$ determines how much energy electrons get from the field during one oscillating period, $eE_0v_F/\omega$, as compared to their average energy $E_F$. Equation (\ref{curr1a}), together with (\ref{integralI-peresform}) and (\ref{funcK}), determines the time dependence of the electric current under the action of an arbitrarily strong ac electric field (\ref{ext-field}).

Equations (\ref{curr1a})--(\ref{funcK}) have been also derived in Ref. \cite{Peres14}, see also \cite{Glazov14} and references therein. Having obtained the expression (\ref{integralI-peresform}), Peres et al. \cite{Peres14} presented it in terms of the hypergeometric function $_2F_1(-\frac12,\frac12;2;\Pi^2)$,
\be 
{\cal I}(\Pi)=\pi \Pi\ _2F_1(-\frac12,\frac12;2;\Pi^2).\label{2F1peres}
\ee
This presentation is valid only at $\Pi\le 1$, i.e., it is unsuitable for the description of the unperturbed solution; for example, at $\Pi>1$ the right-hand side of Eq. (\ref{2F1peres}) is complex while the integral (\ref{integralI-peresform}) is real at all real $\Pi$. Having presented the integral ${\cal I}(\Pi)$ in the form (\ref{2F1peres}) the authors of Ref. \cite{Peres14} further expanded the hypergeometric function in powers of $\Pi\propto {\cal F}_\omega$ up to the fifth order and then used thus expanded (approximate) expression for the integral ${\cal I}(\Pi)$ (\ref{integralI-peresform}) and for the current (\ref{curr1}). This lead them to the prediction of the optical bistability effect in graphene, which is further discussed in Section \ref{bistab?} below.

We proceed in a different way. Our goal is to solve the problem \textit{exactly} not using the Taylor expansion in powers of ${\cal F}_\omega$. To this end, we transform the integral (\ref{integralI-peresform}) as
\be
{\cal I}(\Pi)=
2 \sqrt{1+\Pi^2}\int_{0}^{1}dy \left(\sqrt{1+\Lambda\sqrt{1-y^2}} -\sqrt{1-\Lambda\sqrt{1-y^2}} \right),\label{integralI-myform}
\ee
where 
\be 
\Lambda=\frac{2\Pi}{1+\Pi^2},
\ee
and present the integral (\ref{integralI-myform}) in terms of (another) hypergeometric function,
\be 
{\cal I}(\Pi)=
 \pi    
\frac{\Pi}{\sqrt{1+\Pi^2}}\ _2F_1\left(\frac 14,\frac 34;2;\Lambda^2\right).\label{integralI-myform2}
\ee
The formula (\ref{integralI-myform2}) is valid at $\Lambda\le 1$ and hence, \textit{at any value} of $\Pi\propto {\cal F}_\omega$. 

The corresponding expression for the current then assumes the form
\ba 
j_x(t)&=&
-en_sv_F\int^{\infty}_0 e^{-\xi}  d \xi  
\frac{\Pi}{\sqrt{1+\Pi^2} }\ _2F_1\left(\frac 14,\frac 34,2;\left(\frac{2\Pi}{1+\Pi^2}\right)^2\right)  ;
\ea
it is valid at arbitrary values of the electric field parameter ${\cal F}_\omega$. Rewriting the function $\Pi$, Eq. (\ref{funcK}), as 
\be 
\Pi=- 2{\cal F}_\omega\sin(\omega\tau \xi/2) \sin(\omega t-\omega\tau \xi/2)=-a\sin x,
\ee
where $a=2{\cal F}_\omega\sin(\omega\tau \xi/2)$ and $x=\omega t-\omega\tau \xi/2$, we present the current in the form
\be
j_x(t)=
en_sv_F\int^{\infty}_0 e^{-\xi} 
{\cal M}\Big(2{\cal F}_\omega\sin(\omega\tau \xi/2),\omega t-\omega\tau \xi/2\Big) d \xi  ,\label{curr2}
\ee
where 
\be 
{\cal M }(a,x)=\frac{a\sin x}{\sqrt{1+(a\sin x)^2} } \ 
_2F_1\left(\frac 14,\frac 34,2;\left(\frac{2 a\sin x}{1+(a\sin x)^2}\right)^2\right).
\ee
The function ${\cal M }(a,x)$ is an odd and periodic function of $x$ with the period $2\pi$. It can therefore be expanded in the sine-Fourier series
\be 
{\cal M}(a,x)=\sum_{k=1}^\infty A_k(a)\sin kx,\label{M-expan}
\ee
where 
\be 
A_k(a)=\frac 1\pi\int_{-\pi}^\pi {\cal M }(a,x) \sin kx dx.
\ee
In particular, all functions $A_k(a)$ with an even $k$ equal zero, $A_{2n}(a)=0$, and for $A_k(a)$ with an odd $k$ one can write $A_{2n+1}(a)=aB_{2n+1}(a)$, where
\be 
B_{2n+1}(a)=
\frac 4\pi\int_{0}^{\pi/2} \frac{\sin x\sin(2n+1)x}{\sqrt{1+(a\sin x)^2} } \ 
_2F_1\left(\frac 14,\frac 34,2;\left(\frac{2 a\sin x}{1+(a\sin x)^2}\right)^2\right)  dx .\label{B2n+1}
\ee

The functions $B_{k}(a)$ for $k=1,3,5,7,9$ are shown in Figure \ref{fig:B1}, their properties are discussed in Appendix \ref{app:Bn}. They determine the field dependence of the system response. If the field is small, ${\cal F}_\omega\ll 1$, the parameter $a\propto {\cal F}_\omega$ tends to zero and all $B_{k}(a)$ except $B_{1}(a)$ vanish. Notice that at small $a$ (the low-field limit) the values of $B_k(a)$ fall down very quickly with $k$, $B_k(a)\propto a^{k-1}$; for example, at $a=0.1$ $B_5/B_3=B_3/B_1\approx a^2/32=1/3200$. But at large $a$ (bigger than 1), the functions $B_k(a)$ become quite comparable to each other and fall down slowly with $k$: for example at $a=3$ $B_5/B_3\approx 0.39$ and $B_3/B_1\approx 0.26$. At $a\gg 1$ one gets $B_{2n+1}(a)\approx 4/[\pi a(2n+1)]$, see (\ref{BlargeA}).

\begin{figure}
\includegraphics[width=0.495\textwidth]{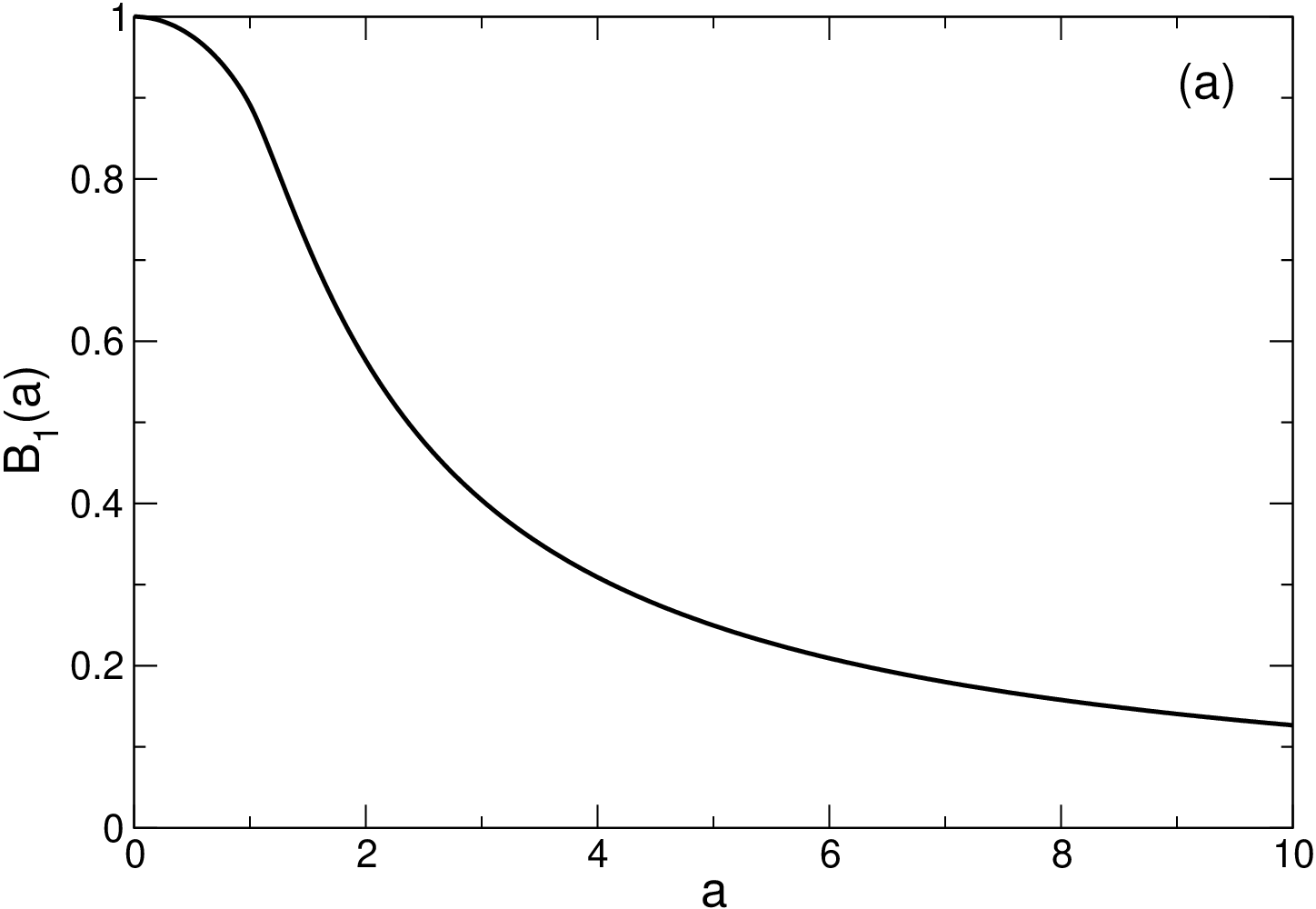}
\includegraphics[width=0.495\textwidth]{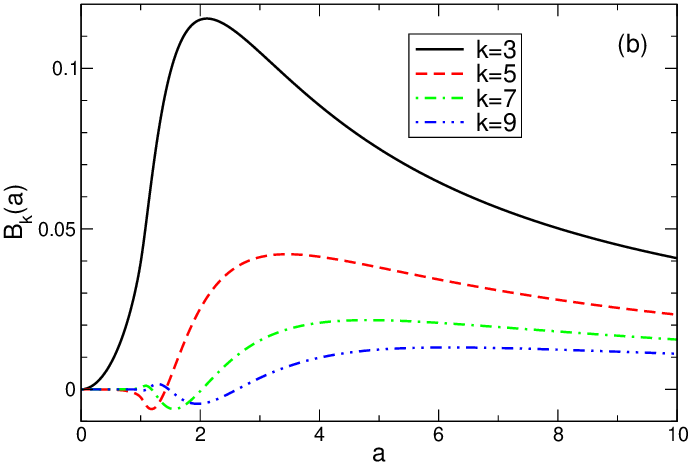}
\caption{The functions $B_k(a)$ for (a) $k=1$ and (b) $k=3,5,7,9$.\label{fig:B1}}
\end{figure}

Using Eq. (\ref{M-expan}) we can now present the current (\ref{curr2}) in the form of the Fourier expansion over odd frequency harmonics: 
\be
j_x(t)= en_sv_F
\sum_{n=0}^\infty \Big(\sin [(2n+1)\omega t] {\cal J}_S^{(2n+1)}
- \cos [(2n+1)\omega t]{\cal J}_C^{(2n+1)}\Big)
  .\label{curr2c}
\ee
The first term in the sum, with 
\be 
{\cal J}_S^{(2n+1)}(\omega\tau,{\cal F}_\tau)={\cal F}_\tau \int^{\infty}_0 e^{-\xi} \frac{\sin(\omega\tau \xi/2)}{\omega\tau/2} B_{2n+1}\left({\cal F}_\tau\frac{\sin(\omega\tau \xi/2)}{\omega\tau/2}\right)
\cos[(2n+1)\omega\tau \xi/2] d \xi,\label{JS}
\ee
represents the current oscillating ``in phase'' with the external electric field (\ref{ext-field}). The second term in (\ref{curr2c}), with 
\be 
{\cal J}_C^{(2n+1)}(\omega\tau,{\cal F}_\tau)={\cal F}_\tau \int^{\infty}_0 e^{-\xi} \frac{\sin(\omega\tau \xi/2)}{\omega\tau/2} B_{2n+1}\left({\cal F}_\tau\frac{\sin(\omega\tau \xi/2)}{\omega\tau/2}\right)
\sin[(2n+1)\omega\tau \xi/2] d \xi,\label{JC}
\ee
represents the current components oscillating ``out of phase'' with the field $E_0\sin\omega t$. In Eqs. (\ref{JS}) and (\ref{JC}) we have introduced a frequency independent field parameter
\be 
{\cal F}_\tau=\frac{eE_0\tau}{\hbar k_F}=\omega\tau {\cal F}_\omega \label{Ftau}
\ee
which is more convenient to use analyzing the frequency dependence of the nonlinear response.

Let us analyze the result (\ref{curr2c}). First, we define an infinite set of generalized ``conductivities'' 
\be 
\sigma_{(2n+1)\omega,\omega}(\omega\tau,{\cal F}_\tau)=
\frac{en_sv_F {\cal F}_\tau}{E_0} \int^{\infty}_0 e^{-\xi} \frac{\sin(\omega\tau \xi/2)}{\omega\tau/2} B_{2n+1}\left({\cal F}_\tau\frac{\sin(\omega\tau \xi/2)}{\omega\tau/2}\right)
e^{i(2n+1)\omega\tau \xi/2} d \xi.\label{generConduct}
\ee
Then Eq. (\ref{curr2c}) assumes the form
\be
j_x(t)= 
\sum_{n=0}^\infty \Big(
\sigma'_{(2n+1)\omega,\omega}
\sin [(2n+1)\omega t] 
- \sigma''_{(2n+1)\omega,\omega}
\cos [(2n+1)\omega t]\Big)E_0,
\label{curr2d}
\ee
similar to the relation (\ref{DrudeParabA}). One sees that thus defined functions $\sigma_{(2n+1)\omega,\omega}$ determine the $(2n+1)\omega$-Fourier components of the current responding to the $\omega$-Fourier component of the external field (\ref{ext-field}). In the low-field limit ${\cal F}_\tau\ll 1$ all functions $\sigma_{(2n+1)\omega,\omega}$, except one, vanish, and one gets the conventional Drude result,
\be 
\sigma_{(2n+1)\omega,\omega}(\omega\tau,{\cal F}_\tau\to 0)=\delta_{n0}\frac{\sigma_0}{1-i\omega\tau},\ \ \ \sigma_0=\frac{n_se^2\tau}{(p_F/v_F)}=\frac{e^2}{\pi\hbar} \frac{E_F\tau}{\hbar},
\ee 
with the mass $m$ replaced by the effective (electron density dependent) mass of graphene electrons at the Fermi level $p_F/v_F=E_F/v_F^2$, compare with (\ref{DrudeConduct}). If the field parameter ${\cal F}_\tau$ is finite, the functions $\sigma_{(2n+1)\omega,\omega}(\omega\tau,{\cal F}_\tau)$, $n>0$, describe the higher (odd) harmonic generation, while the function $\sigma_{\omega,\omega}(\omega\tau,{\cal F}_\tau)$ determines the intensity dependent response of the system at the incident-wave frequency (Kerr effect). Calculating the averaged (over time) energy dissipated in the system due to the scattering, $\langle \bm j\bm E\rangle_t$, 
\be 
\langle \bm j\bm E\rangle_t=\frac 12\sigma'_{\omega,\omega}(\omega\tau,{\cal F}_\tau)
E_0^2,\label{Joule}
\ee 
we see that the Joule heat is determined by the real part of the function $\sigma_{\omega,\omega}(\omega\tau,{\cal F}_\tau)$ only; the higher-harmonic conductivities $\sigma_{(2n+1)\omega,\omega}$ do not contribute to $\langle \bm j\bm E\rangle_t$. 

Let us rewrite now the conductivities $\sigma_{(2n+1)\omega,\omega}$ in the dimensionless form
\be 
\sigma_{(2n+1)\omega,\omega}(\omega\tau,{\cal F}_\tau)=
\sigma_0 {\cal S}_{2n+1}(\omega\tau,{\cal F}_\tau),
\label{sigma(2n+1)ww}
\ee
with
\be 
{\cal S}_{2n+1}(\omega\tau,{\cal F}_\tau)=
\int^{\infty}_0 e^{-\xi} \frac{\sin(\omega\tau \xi/2)}{\omega\tau/2} B_{2n+1}\left({\cal F}_\tau\frac{\sin(\omega\tau \xi/2)}{\omega\tau/2}\right)
e^{i(2n+1)\omega\tau \xi/2} d \xi,
\label{S(2n+1)}
\ee
and analyze the frequency ($\omega\tau$) and field (${\cal F}_\tau$) dependencies of the dimensionless functions ${\cal S}_{2n+1}$. 

\section{Results and their discussion\label{sec:results}}
\subsection{Kerr effect\label{sec:Kerr}}

\begin{figure}
\includegraphics[width=0.495\textwidth]{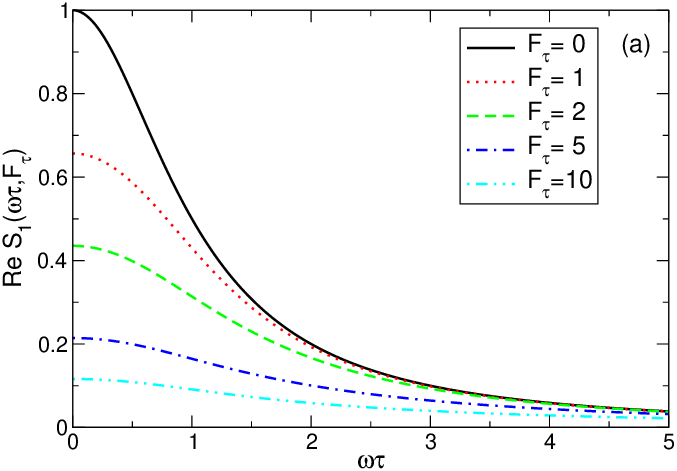}
\includegraphics[width=0.495\textwidth]{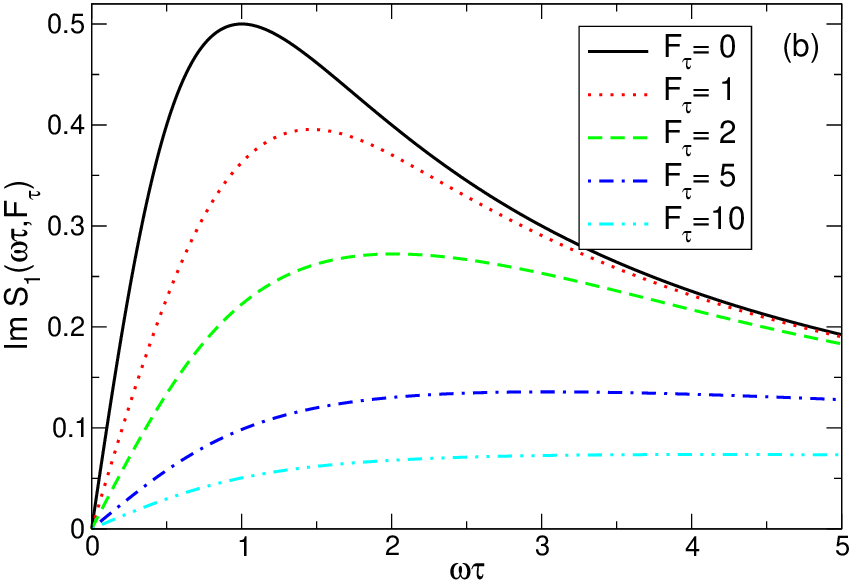}
\caption{\label{fig:S1vsW} The (a) real and (b) imaginary parts of the function $S_1(\omega\tau,{\cal F}_\tau)$ [Eq. (\ref{S(2n+1)}), $n=0$] as a function of $\omega\tau$ at different values of the field parameter ${\cal F}_\tau$.}
\end{figure}

First we consider the function ${\cal S}_{1}(\omega\tau,{\cal F}_\tau)$ which determines the saturable absorption and Kerr effects. Figure \ref{fig:S1vsW} shows the frequency dependence of ${\cal S}_{1}(\omega\tau,{\cal F}_\tau)$ at a few values of the electric field parameter ${\cal F}_\tau$. The curves corresponding to ${\cal F}_\tau=0$ are conventional Drude dependencies, since in the limit ${\cal F}_\tau\to 0$ one gets ${\cal S}_{1}(\omega\tau,0)=1/(1-i\omega\tau)$. When ${\cal F}_\tau$ grows both real and imaginary parts of the conductivity decrease. The suppression of real part ${\cal S}'_{1}(\omega\tau,{\cal F}_\tau)$ at large ${\cal F}_\tau$ corresponds to the saturable absorption. In the limit $\omega\tau\to 0$ one gets 
\be 
S_1(0,{\cal F}_\tau)=
\int^{\infty}_0  \xi e^{-\xi} 
B_1({\cal F}_\tau \xi)d \xi 
 \approx 
\left\{ 
\begin{array}{lr}
1-\frac 34 {\cal F}_\tau^2, & \ {\cal F}_\tau\ll 1 \\
\frac 4{\pi {\cal F}_\tau} , & {\cal F}_\tau\gg 1\\
\end{array}
\right.\label{funcS_W0} .
\ee

\begin{figure}
\includegraphics[width=0.495\textwidth]{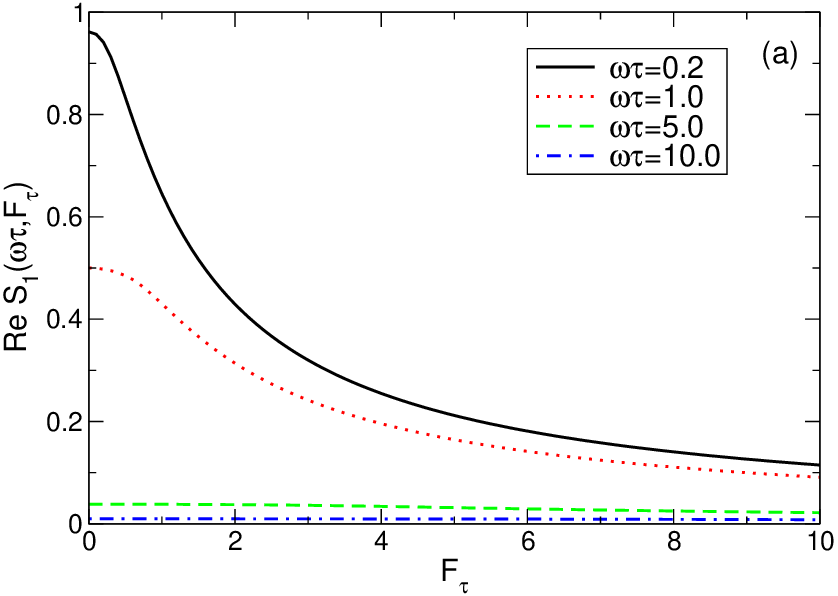}
\includegraphics[width=0.495\textwidth]{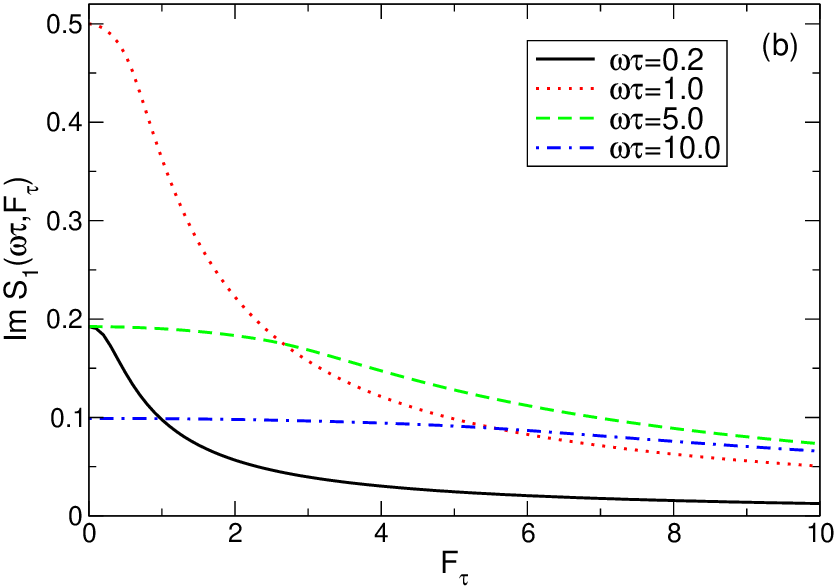}
\caption{\label{fig:S1vsF} The (a) real and (b) imaginary parts of the function $S_1(\omega\tau,{\cal F}_\tau)$ [Eq. (\ref{S(2n+1)}), $n=0$] as a function of ${\cal F}_\tau$ at different values of the frequency parameter $\omega\tau$.}
\end{figure}

Figure \ref{fig:S1vsF} illustrates the field dependence of $S_1(\omega\tau,{\cal F}_\tau)$ at several values of the parameter $\omega\tau$. When the field grows, both the real and imaginary parts of $S_1(\omega\tau,{\cal F}_\tau)$ decrease. This reduction is stronger at small values of $\omega\tau$ and much weaker at $\omega\tau\gg 1$. The field and frequency dependencies of the real part of the conductivity, Figure \ref{fig:S1vsF}a, evidently agree with those recently measured in the experiment of Ref. \cite{Mics15} (see Fig 1c there).

\subsection{Harmonics generation}

\begin{figure}
\includegraphics[width=0.495\textwidth]{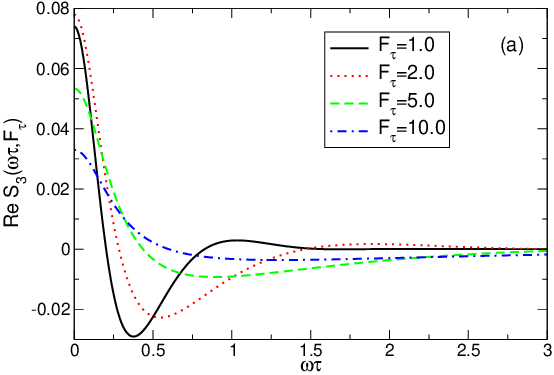}
\includegraphics[width=0.495\textwidth]{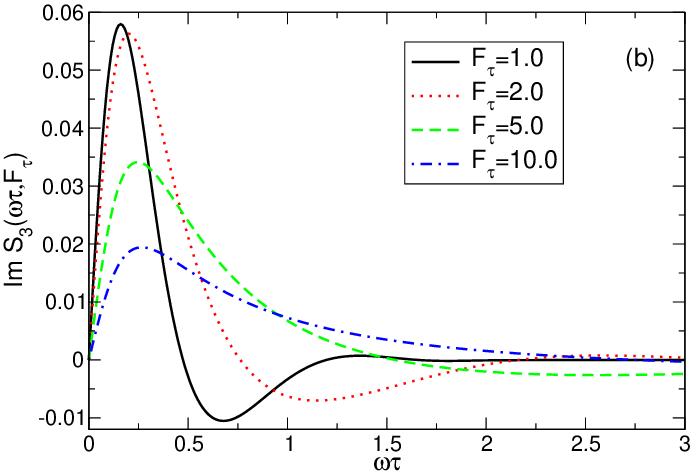}
\caption{\label{fig:cond3WW} The (a) real and (b) imaginary parts of the function $S_{3}(\omega\tau,{\cal F}_\tau)$ [Eq. (\ref{S(2n+1)}), $n=1$] as a function of $\omega\tau$ at different values of the field parameter ${\cal F}_\tau$.}
\end{figure}

\begin{figure}
\includegraphics[width=0.495\textwidth]{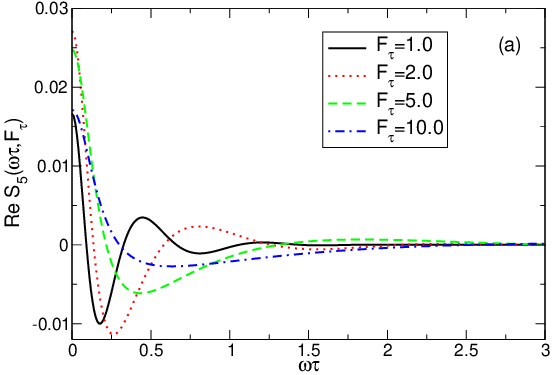}
\includegraphics[width=0.495\textwidth]{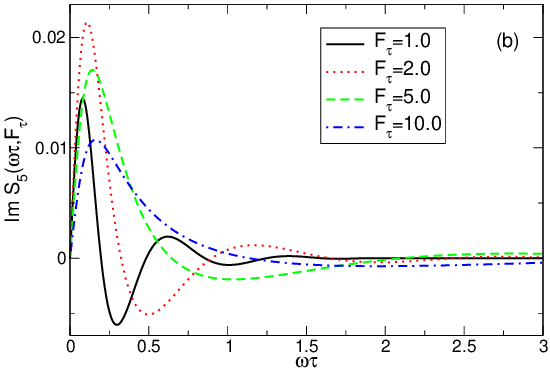}
\caption{\label{fig:cond5WW} The (a) real and (b) imaginary parts of the function $S_{5}(\omega\tau,{\cal F}_\tau)$ [Eq. (\ref{S(2n+1)}), $n=2$] as a function of $\omega\tau$ at different values of the field parameter ${\cal F}_\tau$.}
\end{figure}

\begin{figure}
\includegraphics[width=0.495\textwidth]{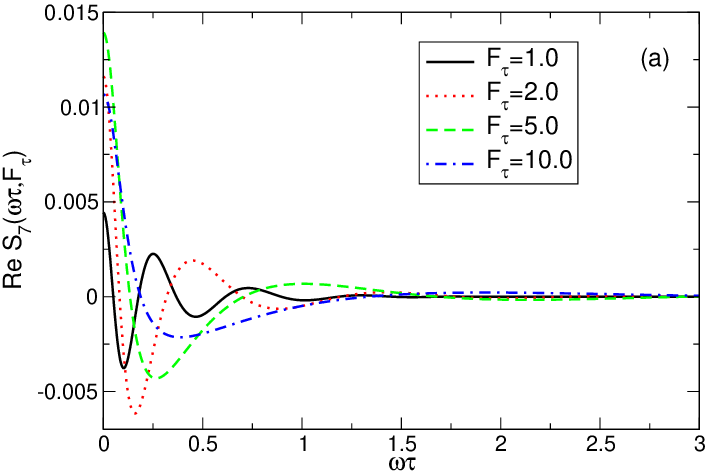}
\includegraphics[width=0.495\textwidth]{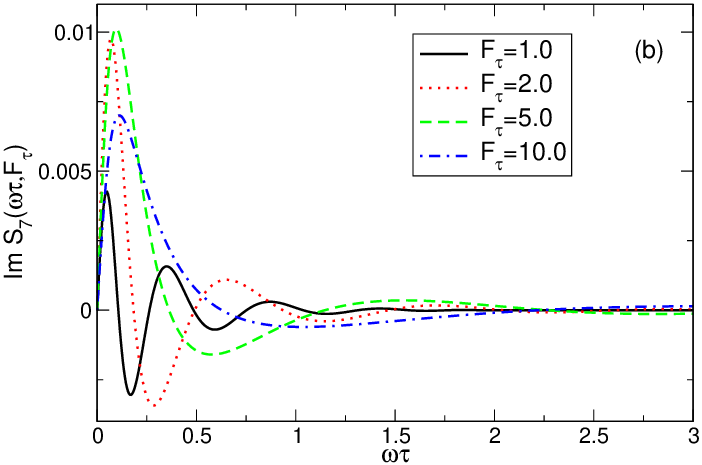}
\caption{\label{fig:cond7WW} The (a) real and (b) imaginary parts of the function $S_{7}(\omega\tau,{\cal F}_\tau)$ [Eq. (\ref{S(2n+1)}), $n=3$] as a function of $\omega\tau$ at different values of the field parameter ${\cal F}_\tau$.}
\end{figure}

Now consider the functions ${\cal S}_{2n+1}(\omega\tau,{\cal F}_\tau)$, $n\ge 1$, responsible for the harmonics generation. Figures \ref{fig:cond3WW} -- \ref{fig:cond7WW} illustrate the frequency dependence of the functions ${\cal S}_{3}(\omega\tau,{\cal F}_\tau)$, ${\cal S}_{5}(\omega\tau,{\cal F}_\tau)$ and ${\cal S}_{7}(\omega\tau,{\cal F}_\tau)$ at a few fixed values of the field parameter ${\cal F}_\tau$. Both real and imaginary parts of ${\cal S}_{2n+1}(\omega\tau,{\cal F}_\tau)$ are quite large at small values of $\omega\tau$ and then quickly decrease (with oscillations) at $\omega\tau \gtrsim 1$. The number of oscillations grows with $n$. 

\begin{figure}
\includegraphics[width=0.495\textwidth]{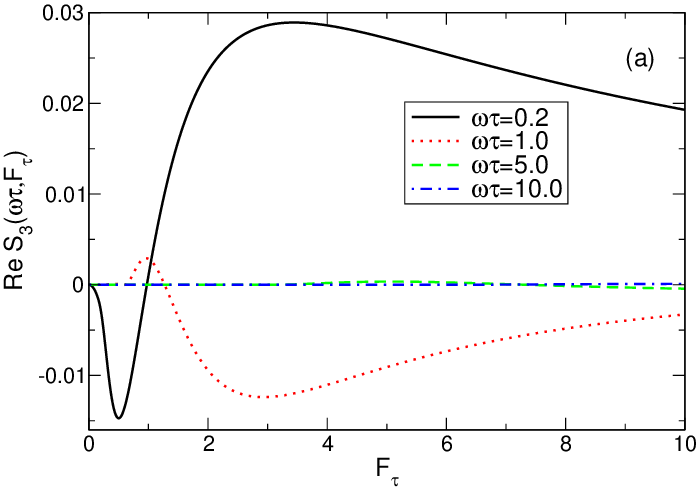}
\includegraphics[width=0.495\textwidth]{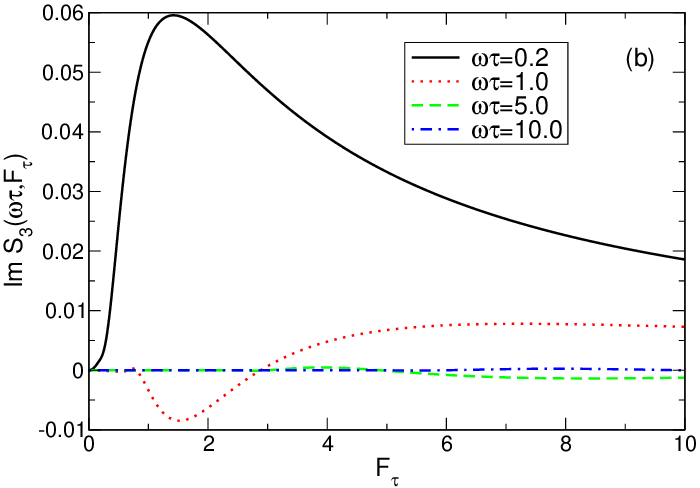}
\caption{\label{fig:S3vsF} The (a) real and (b) imaginary parts of the function $S_3(\omega\tau,{\cal F}_\tau)$ [Eq. (\ref{S(2n+1)}), $n=1$] as a function of ${\cal F}_\tau$ at different values of the frequency parameter $\omega\tau$.}
\end{figure}

\begin{figure}
\includegraphics[width=0.495\textwidth]{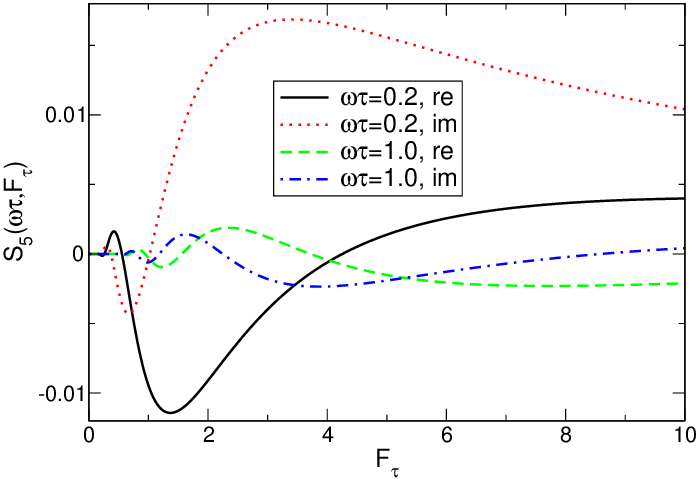}
\caption{\label{fig:S5vsF} The real and imaginary parts of the function $S_5(\omega\tau,{\cal F}_\tau)$ [Eq. (\ref{S(2n+1)}), $n=2$] as a function of ${\cal F}_\tau$ at different values of the frequency parameter $\omega\tau$.}
\end{figure}

\begin{figure}
\includegraphics[width=0.495\textwidth]{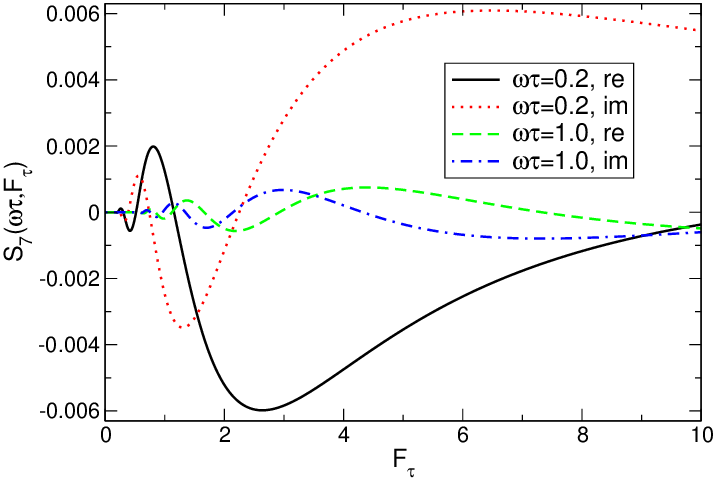}
\caption{\label{fig:S7vsF} The real and imaginary parts of the function $S_7(\omega\tau,{\cal F}_\tau)$ [Eq. (\ref{S(2n+1)}), $n=3$] as a function of ${\cal F}_\tau$ at different values of the frequency parameter $\omega\tau$.}
\end{figure}

\begin{figure}
\includegraphics[width=0.495\textwidth]{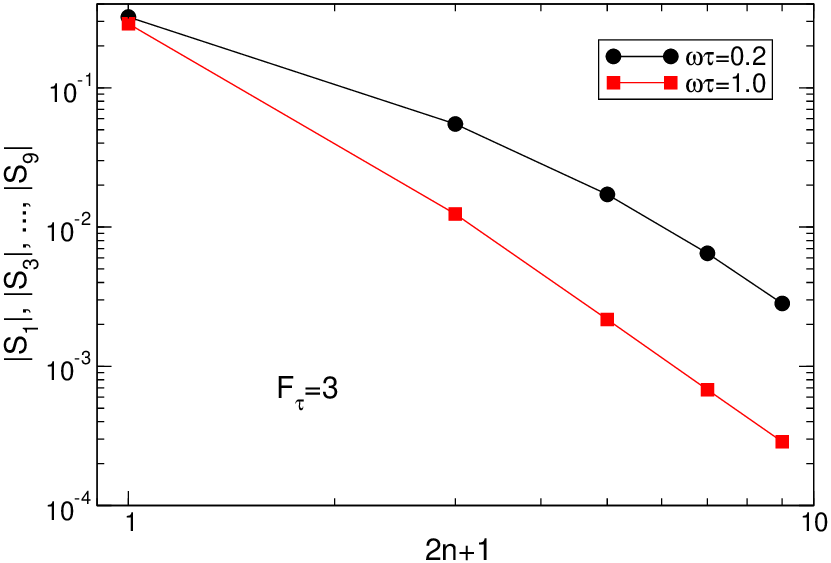}
\caption{\label{fig:SvsN} The absolute values of the functions $|S_{2n+1}(\omega\tau,{\cal F}_\tau)|$, $2n+1=1,3,5,7,9$, as a function of the index $2n+1$ at ${\cal F}_\tau=3$ and different values of the frequency parameter $\omega\tau$.}
\end{figure}

Figures \ref{fig:S3vsF} -- \ref{fig:S7vsF} illustrate the field dependence of the functions ${\cal S}_{3}(\omega\tau,{\cal F}_\tau)$, ${\cal S}_{5}(\omega\tau,{\cal F}_\tau)$ and ${\cal S}_{7}(\omega\tau,{\cal F}_\tau)$ at a few fixed values of the frequency $\omega\tau$. The most interesting features seen on these plots are: (a) the absolute values of the functions ${\cal S}_{2n+1}(\omega\tau,{\cal F}_\tau)$ at $\omega\tau\lesssim 1$ are much larger than those at $\omega\tau\gg 1$, (b) while at small values of the field ${\cal F}_\tau\ll 1$ the values of ${\cal S}_{2n+1}(\omega\tau,{\cal F}_\tau)$ are very small and substantially decrease with the index $n$, at ${\cal F}_\tau\gtrsim 1$ all quantities ${\cal S}_{2n+1}(\omega\tau,{\cal F}_\tau)$ substantially grow and the scale of the functions ${\cal S}_{2n+1}(\omega\tau,{\cal F}_\tau)$ with different $n$ becomes quite comparable with each other. The last finding is very important. It shows that, while the perturbative solutions, obtained at ${\cal F}_\tau\ll 1$, predict very weak amplitudes of the harmonics with $n\gg 1$, the non-perturbative solution shows that at ${\cal F}_\tau\gtrsim 1$ all higher harmonics are quite comparable in their amplitude (this is illustrated in Figure \ref{fig:SvsN}). The crucial condition for observation of higher harmonics at microwave/terahertz frequencies ($\omega\tau\lesssim 1$) is thus ${\cal F}_\tau\gtrsim 1$. The condition ${\cal F}_\tau> 1$ can be rewritten as
\be 
\frac{E_0(\textrm{kV/cm})\textrm{ }\tau(\textrm{ps})}{\sqrt{n_s(10^{12}\textrm{  cm}^{-2})}}>1.16,
\ee
i.e. at $\tau\simeq 1$ ps and the density $\simeq 10^{11}-10^{12}$ cm$^{-2}$ the field of order of one to few kV/cm is already ``strong'' in the sense that ${\cal F}_\tau\gtrsim 1$ and the higher harmonics (3rd, 5th, 7th, 9th) become observable. 

\subsection{Is there an optical bistability in graphene? \label{bistab?}}

As we have mentioned in Section \ref{sec:intro}, an optical bistability effect was predicted in a single graphene layer in Ref. \cite{Peres14}. An electromagnetic wave with the frequency $\omega$ and the amplitude $E_0$ was assumed to be incident on a single isolated graphene layer and the amplitude of the transmitted wave $E_t$ (with the same frequency $\omega$) was calculated. It was shown that the $E_t(E_0)$ dependence is a multivalued function having the $S$-shape.

The result of Ref. \cite{Peres14} was obtained within the perturbation theory. Now we can check, whether the bistability effect survives if the electromagnetic response of graphene is calculated non-perturbatively.

\begin{figure}
\includegraphics[width=0.495\textwidth]{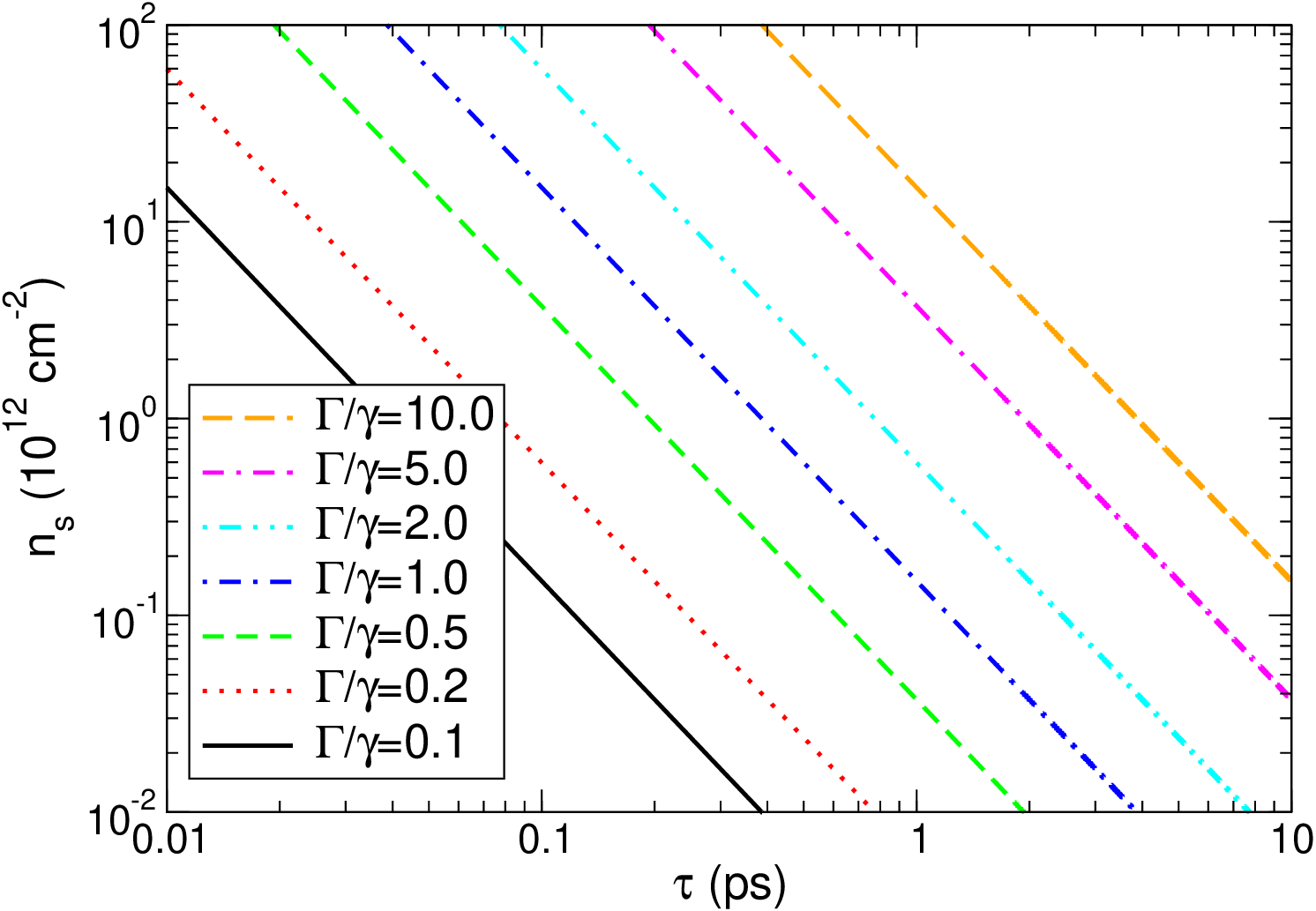}
\caption{\label{fig:Gg}The ratio $\Gamma/\gamma=\Gamma\tau$ of the radiative decay rate $\Gamma$ to the scattering relaxation rate $\gamma$ at different values of $\tau$ (in ps) and of the electron density $n_s$ (in units of $10^{12}$ cm$^{-2}$).
}
\end{figure}

Similar to Ref. \cite{Peres14}, we consider the incidence of radiation on a single graphene layer placed at the plane $z=0$. The electric and magnetic fields of the wave are then 
\be 
E_x(t)=E_0^\omega e^{-i\omega t}\left\{
\begin{array}{ll}
e^{i\omega z/c}+\rho_a e^{-i\omega z/c}, & z<0 \\
\tau_a e^{i\omega z/c}, & z>0 \\
\end{array}
\right. ,
\ee
\be 
H_y(t)=E_0^\omega e^{-i\omega t}\left\{
\begin{array}{ll}
e^{i\omega z/c}-\rho_a e^{-i\omega z/c}, &  z<0\\
\tau_a e^{i\omega  z/c},& z>0
\end{array}
\right.
\ee
where $\rho_a$ and $\tau_a$ are complex reflection and transmission amplitudes of the wave, and the \textit{complex} amplitude of the field $E_0^\omega$ is twice as big as the \textit{real} amplitude (as in Eq. (\ref{ext-field})), $|E_0^\omega|=2E_0$. Applying the conventional boundary conditions at the plane $z=0$ we obtain the transmission ($\tau_a$) and reflection ($\rho_a$) amplitudes,
\be 
\tau_a=1+\rho_a=\frac 1{1+\frac {2\pi }{c}\sigma_{\omega,\omega}(\omega\tau,{\cal F}_{\tau,z= 0})},
\ee
as well as the relation between the field at the plane $z=0$ and the field of the incident wave,
\be  
E_{z=0}^\omega=\tau_a E_0^\omega .
\ee
Here the field-dependent complex conductivity $\sigma_{\omega,\omega}(\omega\tau,{\cal F}_{\tau })$ is defined in Eq. (\ref{generConduct}) and the field parameter ${\cal F}_{\tau,z= 0}$ is determined by the electric field $E_{z=0}^\omega$ \textit{at the plane} $z=0$. 
The relation between the electric field parameters ${\cal F}_{\tau,z= 0}$ and ${\cal F}_{\tau 0}$ (the latter one is proportional to the absolute value of the incident-wave field, see Eq. (\ref{Ftau})) then assumes the form 
\be
{\cal F}_{\tau,z=0}=|\tau_a| {\cal F}_{\tau 0} =\frac{{\cal F}_{\tau 0}}{\left|1 +\frac {2\pi }{c}\sigma_{\omega,\omega}(\omega\tau ,{\cal F}_{\tau,z=0})\right|}.
\ee
This can be rewritten as
\be
{\cal F}_{\tau,z=0}\left|1 +\Gamma\tau {\cal S}_1(\omega\tau ,{\cal F}_{\tau,z=0})\right|={\cal F}_{\tau 0},\label{Frelation}
\ee
where 
\be 
\Gamma=2
\frac{e^2}{\hbar c} \frac{E_F}{\hbar}=2
\frac{e^2}{\hbar c} v_F\sqrt{\pi n_s}
\ee
is the radiative decay rate first derived in Ref. \cite{Mikhailov08a} (see Eq. (24) there). Figure \ref{fig:Gg} shows the ratio of the radiative decay rate to the scattering relaxation rate $\Gamma\tau=\Gamma/\gamma$ at different values of the electron density $n_s$ and the scattering time $\tau$. The larger is the electron density and their mobility, the bigger is the parameter $\Gamma\tau$.

\begin{figure}
\includegraphics[width=0.495\textwidth]{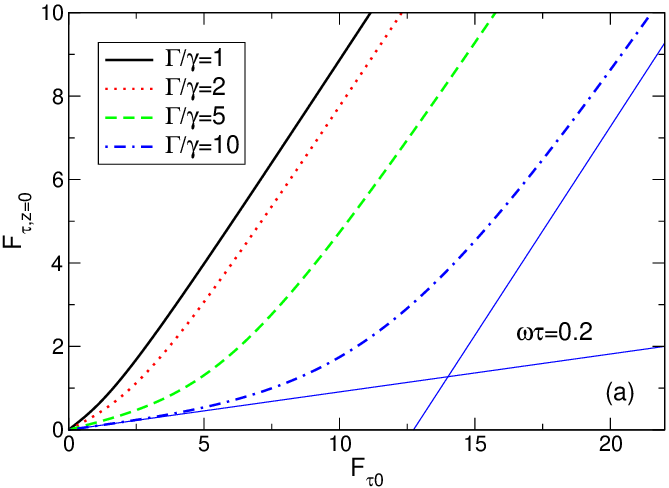}
\includegraphics[width=0.495\textwidth]{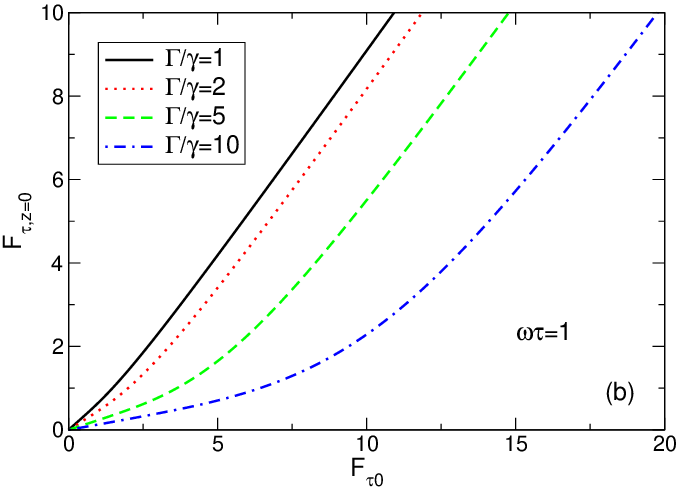}
\includegraphics[width=0.495\textwidth]{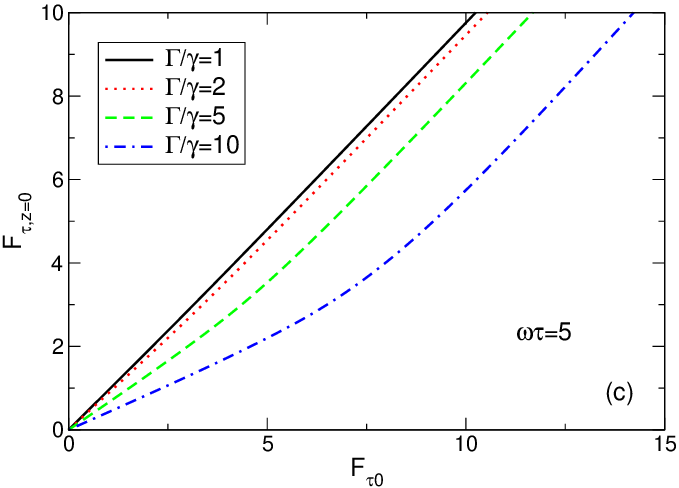}
\caption{\label{fig:fields}The field parameter ${\cal F}_{\tau,z= 0}$ related to the electric field at the plane $z=0$ as a function of the field parameter ${\cal F}_{\tau 0}$ related to the electric field of the incident wave at (a) $\omega\tau=0.2$, (b) $\omega\tau=1$ and (c) $\omega\tau=5$ and at a few different values of the ratio $\Gamma/\gamma$. The thin blue lines in (a) show the asymptotes (\ref{as1}) and (\ref{as2}) to the curve with $\Gamma/\gamma=10$. }
\end{figure}

Now consider the relation (\ref{Frelation}) between the field parameters ${\cal F}_{\tau 0}$ (proportional to the electric field of the incident wave) and ${\cal F}_{\tau,z= 0}$ (proportional to the electric field at the plane $z=0$).  The dependence ${\cal F}_{\tau,z= 0}({\cal F}_{\tau 0})$ is shown in Figure \ref{fig:fields}. One sees that, if $\Gamma/\gamma\lesssim 1$, the difference between ${\cal F}_{\tau,z= 0}$ and ${\cal F}_{\tau 0}$ is very small. If $\Gamma/\gamma\gg 1$, such a difference does exist: at small values of the field the parameter ${\cal F}_{\tau,z= 0}$ grows substantially slower than ${\cal F}_{\tau 0}$; when the field gets stronger, the parameter ${\cal F}_{\tau,z= 0}$ grows together with ${\cal F}_{\tau 0}$ with the slope $d{\cal F}_{\tau,z= 0}/d{\cal F}_{\tau 0}\simeq 1$. The difference between the field at $z=0$ and the incident-wave field is more pronounced at $\omega\tau\ll 1$. This behavior can be understood by analyzing the formulas (\ref{Frelation}) and (\ref{funcS_W0}) [the expansion (\ref{funcS_W0}) is valid at $\omega\tau\ll 1$]. At ${\cal F}_{\tau,z= 0}\ll 1$ the function ${\cal S}_1\approx 1$ and we get from (\ref{Frelation})
\be
{\cal F}_{\tau,z=0}\approx \frac {{\cal F}_{\tau 0}}{1 +\Gamma/\gamma},\ \ \ \omega\tau\ll 1,\  {\cal F}_{\tau,z= 0}\ll 1;\label{as1}
\ee
if ${\cal F}_{\tau,z= 0}\gg 1$ the function ${\cal S}_1\approx 4/\pi {\cal F}_{\tau,z= 0}$ and we get 
\be
{\cal F}_{\tau,z=0}  ={\cal F}_{\tau 0}-\frac{4\Gamma}{\pi\gamma},\ \ \ \omega\tau\ll 1,\ {\cal F}_{\tau,z= 0}\gg 1.\label{as2}
\ee
The asymptotes (\ref{as1}) and (\ref{as2}) (to the curve with $\Gamma/\gamma=10$) are shown in Figure \ref{fig:fields}(a) by thin blue lines.

\begin{figure}
\includegraphics[width=0.495\textwidth]{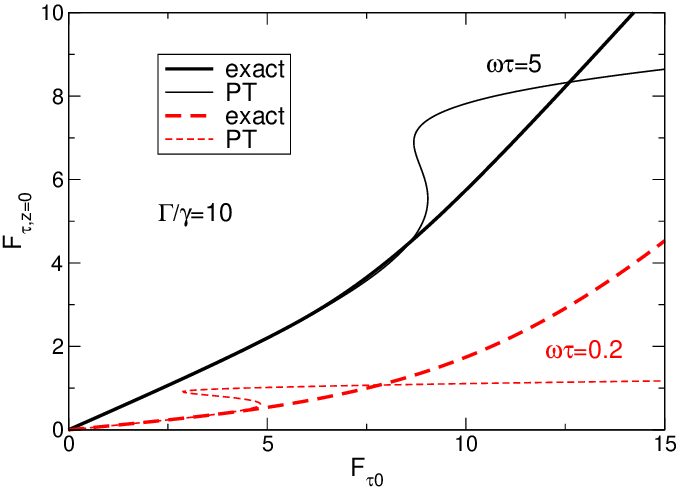}
\caption{\label{fig:fieldsApprox}The field parameter ${\cal F}_{\tau,z= 0}$ as a function of ${\cal F}_{\tau 0}$ at $\Gamma/\gamma=10$ and two (small and large) values of $\omega\tau$, calculated within the perturbation theory (thin curves, labeled PT), as it was done in Ref. \cite{Peres14}, and non-perturbatively (thick curves, this work). One sees that the ``bistability effect'' disappears if the problem is solved exactly. }
\end{figure}

The ``optical bistability'' effect predicted in Ref. \cite{Peres14} is not seen in Figure \ref{fig:fields} at all (small and large) values of both parameters $\omega\tau$ and $\Gamma/\gamma$. Since in Ref. \cite{Peres14} the problem was solved within the perturbation theory, by expanding the conductivity in powers of the electric field up to $O(E^4)$, it is reasonable to assume that the reason of the disagreement is related to the use of this expansion. To check this, we have expanded the function ${\cal S}_1(\omega\tau ,{\cal F}_{\tau,z=0})$ in powers of the electric field up to the same order as it was done in Ref. \cite{Peres14} (see Appendix \ref{app:Bn}) and obtained the result shown in Figure \ref{fig:fieldsApprox}. The perturbation-theory solution does lead to the multi-valued, $S$-shaped dependencies of the output-input characteristics (as seen below, the intensity of the transmitted and the incident light is proportional to $|{\cal F}_{\tau,z= 0}|^2$ and $|{\cal F}_{\tau 0}|^2$, respectively), but this ``bistability'' disappears if the problem is solved exactly. 

\subsection{Scattering of radiation at an isolated graphene layer. Power induced transparency enhancement
} 

To complete the study let us now consider the reflection, transmission and absorption coefficients of the wave incident on an isolated graphene layer (in this Section we consider only the response of graphene at the frequency $\omega$ neglecting the harmonics generation effect). To this end we introduce the intensity of the incident wave ${\cal J}_{i}$ related to the \textit{real} amplitude $E_0$ of the electric field of the incident wave,
\be 
{\cal J}_{i}=\frac c{8\pi}E_0^2,\label{Ji}
\ee
the intensities of the transmitted ${\cal J}_t$ and reflected ${\cal J}_r$ waves,  
\be 
{\cal J}_t=\frac {c}{8\pi}E_0^2\left|\tau_a\right|^2,\label{Jt}
\ee
\be 
{\cal J}_r=\frac {c}{8\pi}E_0^2\left|\rho_a\right|^2,\label{Jr}
\ee
and a characteristic intensity value 
\be 
{\cal J}_0\equiv \frac{c}{8\pi}\left(\frac{\hbar k_F}{e\tau}\right)^2 =\frac{\hbar c}{8e^2}\frac{\hbar  n_s}{\tau^2} \approx 1.8 \times \frac{n_s[10^{12}\mathrm{ cm}^{-2}]}{[\tau\mathrm{ (ps)}]^2} \ \frac{\mathrm {kW}}{\mathrm{cm}^2};\label{J0}
\ee
using (\ref{J0}) the field parameters ${\cal F}_{\tau 0}$ and ${\cal F}_{\tau,z= 0}$ can be written as the ratio of the incident and transmitted waves intencities to ${\cal J}_0$:
\be 
{\cal F}_{\tau 0}^2=\frac{{\cal J}_i}{{\cal J}_0}, \ \ \ {\cal F}_{\tau,z= 0}^2=\frac{{\cal J}_t}{{\cal J}_0}.\label{FrelatedJ}
\ee
The transmission ${\cal T}$, reflection ${\cal R}$ and absorption ${\cal A}$ coefficients then assume the form
\be
{\cal T}=\frac{{\cal J}_t}{{\cal J}_i}=\frac{1}{\left|1 +\frac {2\pi }{c}\sigma_{\omega,\omega}(\omega\tau ,{\cal F}_{\tau,z=0})\right|^2}=\frac{1}{\left|1 +\frac {\Gamma }{\gamma}{\cal S}_1(\omega\tau ,{\cal F}_{\tau,z=0})\right|^2},\label{T}
\ee
\be 
{\cal R}=\frac{{\cal J}_r}{{\cal J}_i}=\frac{\left|\frac {\Gamma }{\gamma}{\cal S}_1(\omega\tau ,{\cal F}_{\tau,z=0})\right|^2}{\left|1 +\frac {\Gamma }{\gamma}{\cal S}_1(\omega\tau ,{\cal F}_{\tau,z=0})\right|^2},\label{R}
\ee
\be 
{\cal A}
=\frac{ 2\frac {\Gamma }{\gamma}{\cal S}'_1(\omega\tau ,{\cal F}_{\tau,z=0})}{\left|1+
\frac {\Gamma }{\gamma}{\cal S}_1(\omega\tau ,{\cal F}_{\tau,z=0})\right|^2}.\label{A}
\ee
The field factor ${\cal F}_{\tau,z=0}$ in Eqs. (\ref{T})--(\ref{A}) depends on ${\cal J}_t$, ${\cal F}_{\tau,z=0}=\sqrt{{\cal J}_t/{\cal J}_0}$, Eq. (\ref{FrelatedJ}), therefore, in order to find the required dependencies of ${\cal T}$, ${\cal R}$ and ${\cal A}$ on the incident wave intensity ${\cal J}_i$, one should first find the relation between ${\cal J}_t$ and ${\cal J}_i$ from Eq. (\ref{T}) and then substitute the thus found intensity ${\cal J}_t({\cal J}_i)$ in Eqs. (\ref{R}) and (\ref{A}). The functions ${\cal T}$, ${\cal R}$, and ${\cal A}$ in Eqs. (\ref{T})--(\ref{A}) depend on three dimensionless parameters: $\omega\tau$, $\Gamma /\gamma$, and ${\cal J}_i/{\cal J}_0$.

\begin{figure}
\includegraphics[width=0.495\textwidth]{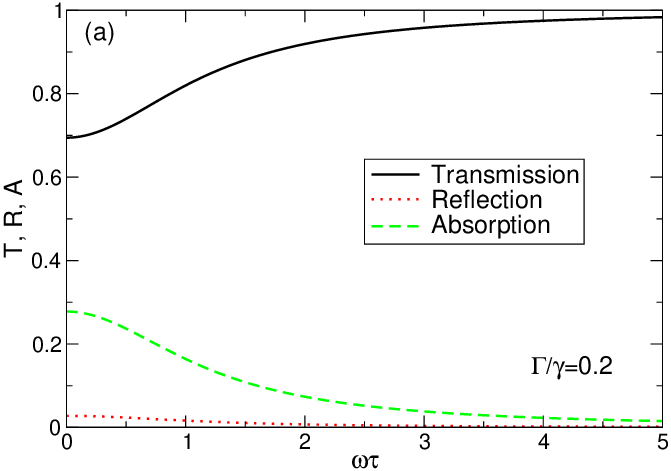}
\includegraphics[width=0.495\textwidth]{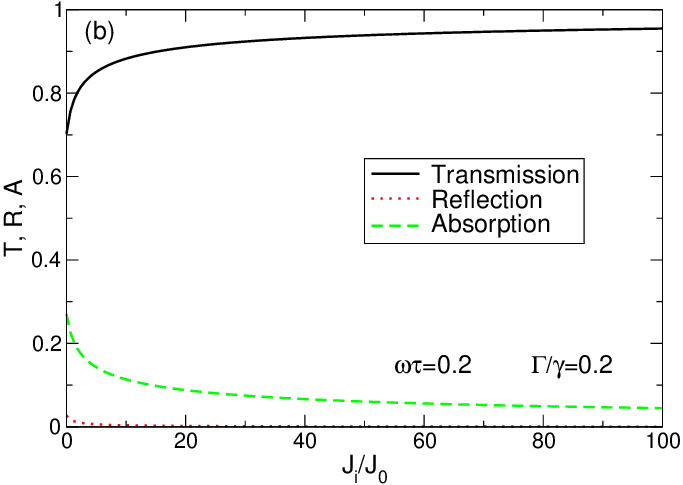}
\caption{\label{fig:TRA_Gg02}The transmission, reflection and absorption coefficients as a function of (a) the frequency parameter $\omega\tau$ at $\Gamma/\gamma=0.2$ and low intensity of the incident wave ${\cal J}_i\to 0$ (the linear response regime), and (b) of the intensity of the incident wave ${\cal J}_i/{\cal J}_0$  at $\omega\tau=0.2$. At larger values of $\omega\tau$ the behavior of ${\cal T}$, ${\cal R}$, and ${\cal A}$ is similar.
}
\end{figure}

\begin{figure}
\includegraphics[width=0.495\textwidth]{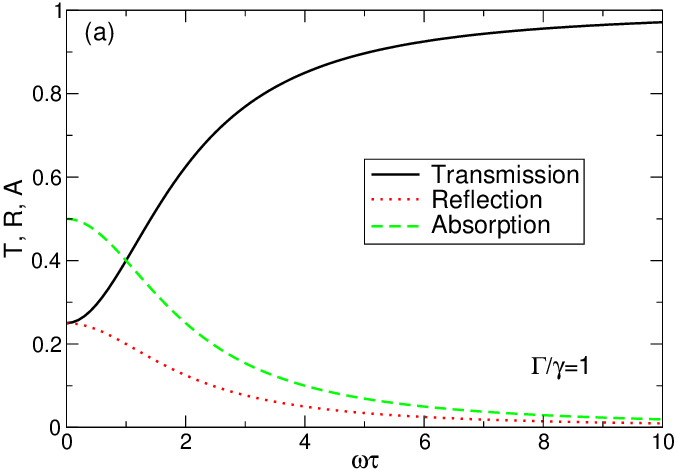}
\includegraphics[width=0.495\textwidth]{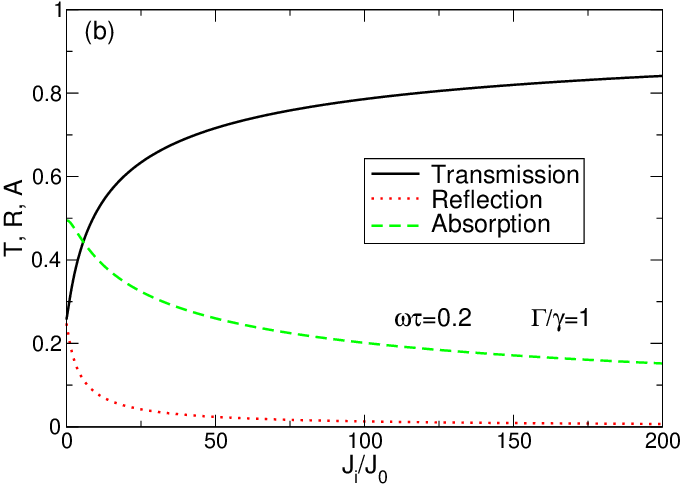}
\includegraphics[width=0.495\textwidth]{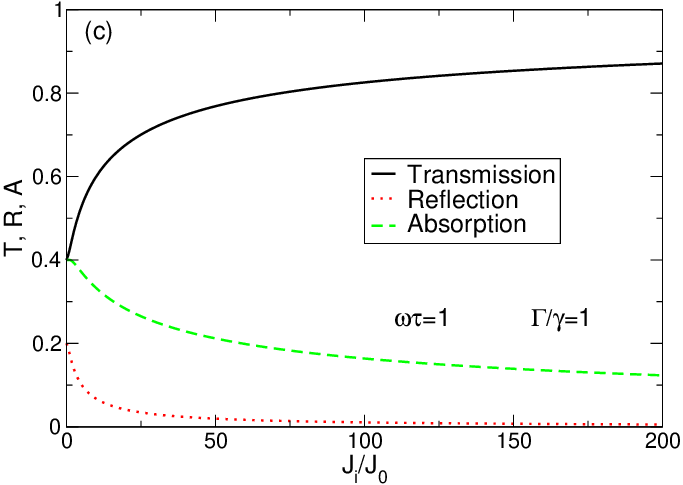}
\includegraphics[width=0.495\textwidth]{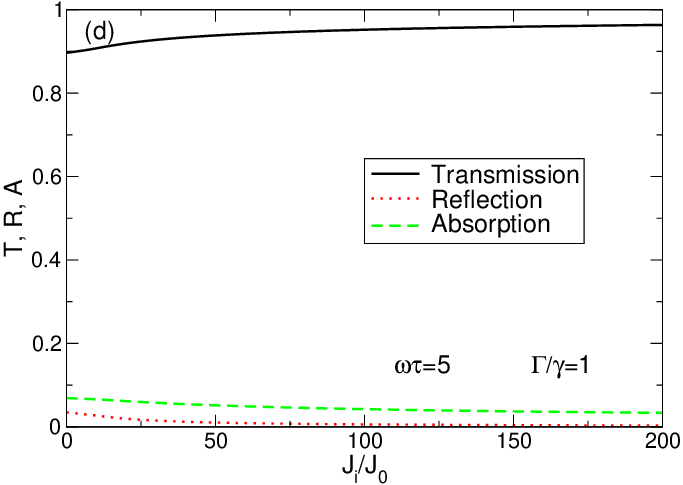}
\caption{\label{fig:TRA_Gg1}The transmission, reflection and absorption coefficients as a function of (a) the frequency parameter $\omega\tau$ at $\Gamma/\gamma=1$ and low intensity of the incident wave ${\cal J}_i\to 0$ (the linear response regime), and of the intensity of the incident wave ${\cal J}_i/{\cal J}_0$  at (b) $\omega\tau=0.2$, (c) $\omega\tau=1$, and (d) $\omega\tau=5$.
}
\end{figure}

\begin{figure}
\includegraphics[width=0.495\textwidth]{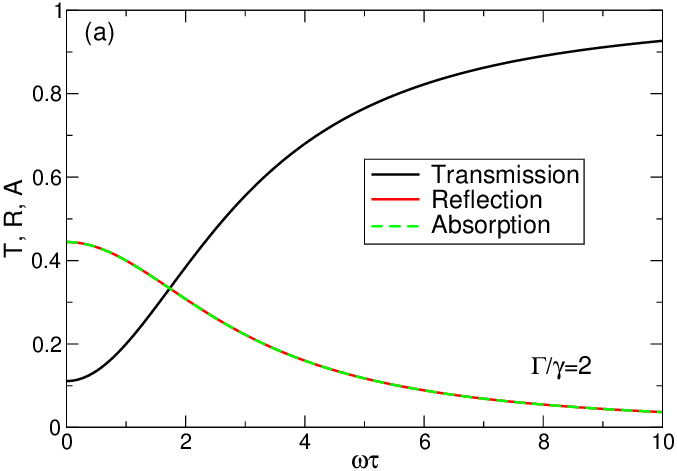}
\includegraphics[width=0.495\textwidth]{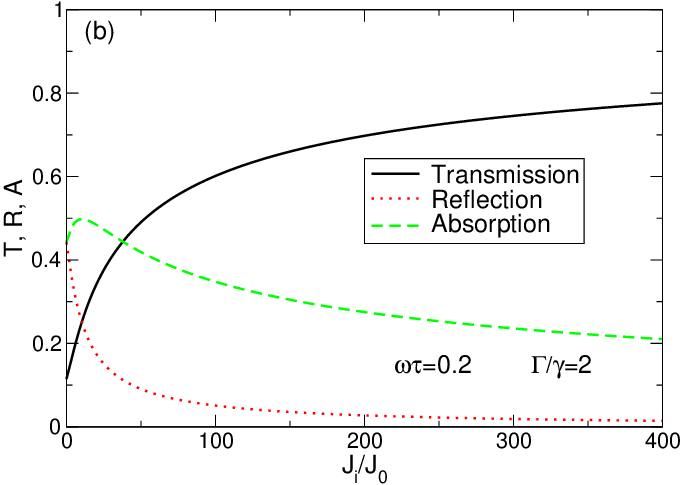}
\includegraphics[width=0.495\textwidth]{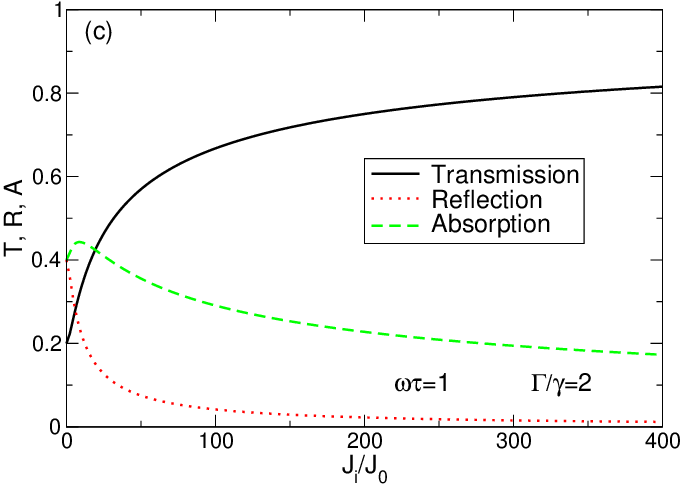}
\includegraphics[width=0.495\textwidth]{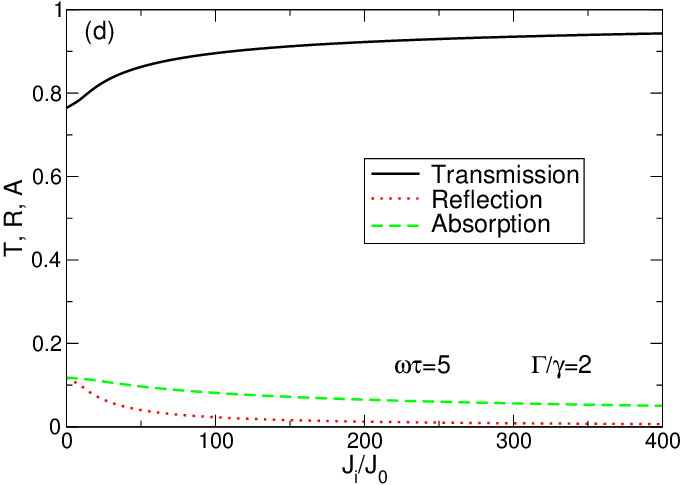}
\caption{\label{fig:TRA_Gg2}The transmission, reflection and absorption coefficients as a function of (a) the frequency parameter $\omega\tau$ at $\Gamma/\gamma=2$ and low intensity of the incident wave ${\cal J}_i\to 0$ (the linear response regime), and of the intensity of the incident wave ${\cal J}_i/{\cal J}_0$  at (b) $\omega\tau=0.2$, (c) $\omega\tau=1$, and (d) $\omega\tau=5$.
}
\end{figure}

\begin{figure}
\includegraphics[width=0.495\textwidth]{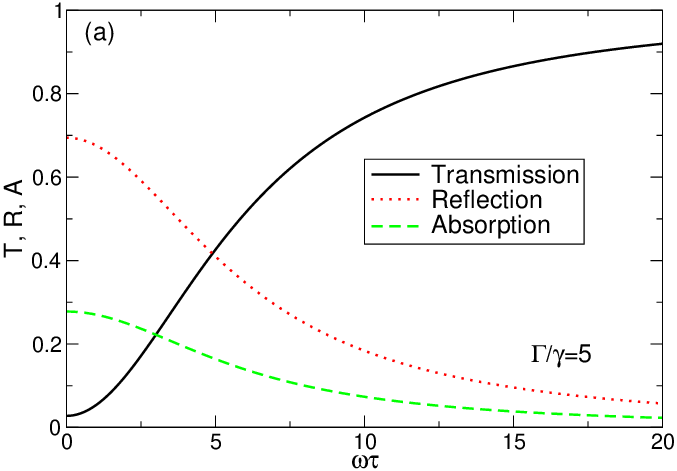}
\includegraphics[width=0.495\textwidth]{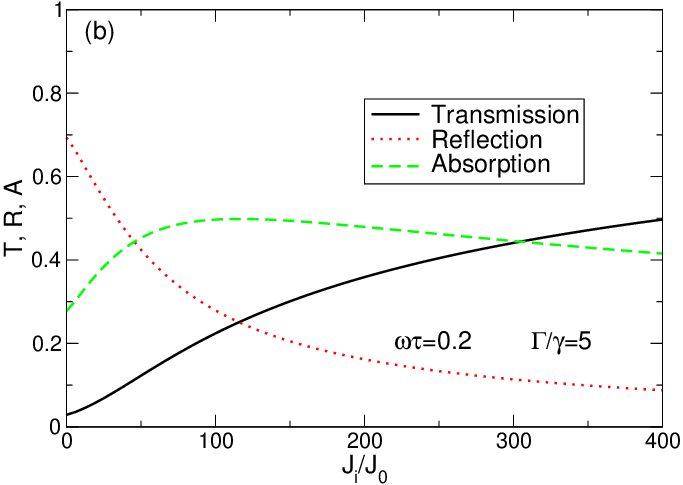}
\includegraphics[width=0.495\textwidth]{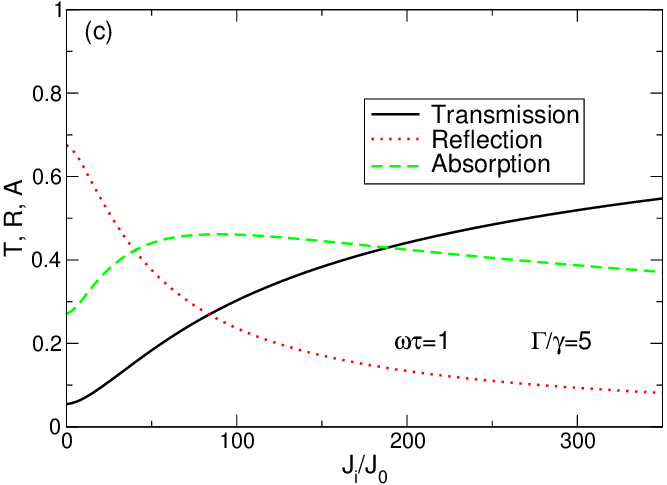}
\includegraphics[width=0.495\textwidth]{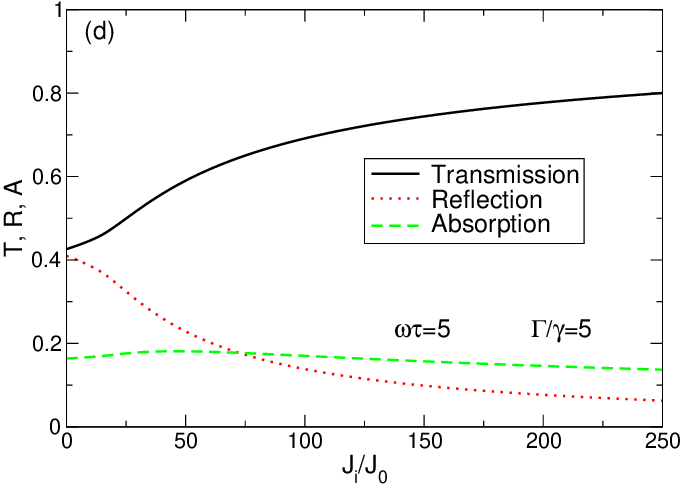}
\caption{\label{fig:TRA_Gg5}The transmission, reflection and absorption coefficients as a function of (a) the frequency parameter $\omega\tau$ at $\Gamma/\gamma=5$ and low intensity of the incident wave ${\cal J}_i\to 0$ (the linear response regime), and of the intensity of the incident wave ${\cal J}_i/{\cal J}_0$  at (b) $\omega\tau=0.2$, (c) $\omega\tau=1$, and (d) $\omega\tau=5$.
}
\end{figure}

First, consider the linear response limit ${\cal J}_i/{\cal J}_0\ll 1$. Then ${\cal F}_{\tau 0}$ and ${\cal F}_{\tau,z= 0}$ are small as compared to unity and ${\cal S}_1(\omega\tau ,{\cal F}_{\tau,z=0}\to 0)\approx 1/(1-i\omega\tau)$. The ${\cal T}$, ${\cal R}$, and ${\cal A}$ coefficients are then
\be
{\cal T}=\frac{\gamma^2+\omega^2}{(\gamma+\Gamma)^2+\omega^2}=
1-\frac{\Gamma^2+2\gamma\Gamma}{(\gamma+\Gamma)^2+\omega^2},\ \ \ {\cal J}_i/{\cal J}_0\ll 1,\label{T0}
\ee
\be 
{\cal R}=\frac{\Gamma^2 }{(\gamma+\Gamma)^2 +\omega^2},\ \ \ {\cal J}_i/{\cal J}_0\ll 1,\label{R0}
\ee
\be 
{\cal A}
=\frac{ 2\Gamma \gamma}{(\gamma+\Gamma)^2+\omega^2}, \ \ \ {\cal J}_i/{\cal J}_0\ll 1.\label{A0}
\ee
The linear-response dependencies (\ref{T0})--(\ref{A0}) are shown in Figures \ref{fig:TRA_Gg02}(a)--\ref{fig:TRA_Gg5}(a). One sees that there exist four physically different regimes. If the radiative decay rate is smaller than the scattering rate, $\Gamma\ll\gamma$, Fig. \ref{fig:TRA_Gg02}, the energy of the incident wave is mainly dissipated into the lattice due to the scattering of electrons by phonons and impurities. The reflection coefficient (red dotted curves) is much smaller than the absorption coefficient (green dashed curves) at all frequencies, and the transmission coefficient is approximately equal to one minus absorption coefficient. At low frequencies ($\omega\tau\ll 1$) ${\cal A}$ is smaller than 50\% in this regime (much smaller if $\Gamma\ll\gamma$), while the transmission is larger than 50\% (much larger if $\Gamma\ll\gamma$).

If the incident wave intensity grows, Figure \ref{fig:TRA_Gg02}(b), the absorption and reflection coefficient fall down while the transmission one substantially grows [in the example of Figure \ref{fig:TRA_Gg02}(b), $\Gamma/\gamma=0.2$, from 70\% at ${\cal J}_i/{\cal J}_0\to 0$ up to 95\% at ${\cal J}_i/{\cal J}_0=100$ and 97.6\% at ${\cal J}_i/{\cal J}_0=400$]. Figure \ref{fig:TRA_Gg02}(b) shows the ${\cal J}_i$ dependencies of the ${\cal TRA}$ coefficients at $\omega\tau=0.2$; at other frequencies these dependencies are qualitatively the same.

Another interesting case is realized at $\Gamma/\gamma=1$, Figure \ref{fig:TRA_Gg1}. In this situation, in the linear-response regime and at low frequencies, $\omega\tau\ll 1$, the radiative and dissipative losses are equal, the system absorbs 50\% of the incident radiation (this corresponds to the matched load regime), and the transmission and reflection coefficients equal 25\% each, Figure \ref{fig:TRA_Gg1}(a). When the intensity of the incident wave increases, the reflection and absorption coefficients dramatically decrease, while the transmission coefficient substantially grows, Figures \ref{fig:TRA_Gg1}(b)-(d). For example, at $\omega\tau=0.2$ and ${\cal J}_i/{\cal J}_0=200$, Figure \ref{fig:TRA_Gg1}(b), the system absorbs about 15\%, reflects less than 0.7\% and transmits more than 84\%. 

The third interesting case is analyzed in Figure \ref{fig:TRA_Gg2}. Here $\Gamma=2\gamma$ and the linear-response reflection and absorption coefficients are equal at all frequencies (at least in the Drude model), Figure \ref{fig:TRA_Gg2}(a). At $\omega\tau\ll 1$ the graphene layer absorbs and reflects about 44.4\% and transmits $\sim 11.1$\%. When the incident wave intensity grows, the behavior of the ${\cal TRA}$-coefficients differs from that one in the two previous cases. The reflection coefficient quickly falls down as before (red dotted curves in Figures \ref{fig:TRA_Gg2}(b)--(d)) reaching the values below 1-2\% at ${\cal J}_i/{\cal J}_0\gtrsim 200$. But the behavior of ${\cal A}$ is different. At small values of ${\cal J}_i/{\cal J}_0$ it first grows up reaching at small frequencies $\sim 50$\% and only after that falls down. This growing effect is especially pronounced at $\omega\tau\lesssim 1$, Figs. \ref{fig:TRA_Gg2}(b)--(c). The transmission coefficient continuously grows with ${\cal J}_i/{\cal J}_0$ at all frequencies reaching $\sim 80$\% at $\omega\tau\lesssim 1$ and ${\cal J}_i/{\cal J}_0\sim 400$, Figs. \ref{fig:TRA_Gg2}(b)--(c), and even $\sim 95$\% at higher frequencies, Fig. \ref{fig:TRA_Gg2}(d).

Finally, Figure \ref{fig:TRA_Gg5} shows the transmission, reflection and absorption coefficients at $\Gamma\gg\gamma$, when the radiative losses substantially exceed the dissipative losses ($\Gamma/\gamma=5$ in this Figure). Such a situation is realized when either the density or the mobility of the electrons (or both) are large. In this case the transmission coefficient of the 2D layer, in the linear response regime ${\cal J}_i\to 0$, is very small at low frequencies $\omega\tau\ll 1$, see Figure \ref{fig:TRA_Gg5}(a). The energy of the incident wave is either reflected or absorbed, with ${\cal R}/{\cal A}=\Gamma/2\gamma\gg 1$. In the illustrative example of Figure \ref{fig:TRA_Gg5}  ($\Gamma/\gamma=5$) the layer transmits, at low frequencies, less than 3\% and reflects almost 70\%. 
  
When the intensity ${\cal J}_i$ increases, the absorption coefficient first grows up, from $\sim 28$\% up to 50\% at $\omega\tau\ll 1$, in the example of Figure \ref{fig:TRA_Gg5}(b), and then falls down at ${\cal J}_i\to\infty$. The reflection coefficient falls down with the growing ${\cal J}_i$ as in all considered cases. The transmission coefficient continuously increases with ${\cal J}_i$, from just a few percent up to 50--80\% dependent on the frequency and the intensity ${\cal J}_i$.

Thus, in all considered cases the reflection coefficient strongly decreases while the transmission coefficient strongly increases with the growing intensity ${\cal J}_i$. The absorption coefficient either monotonously falls down with ${\cal J}_i$ at small values of $\Gamma/\gamma$, or first grows and then decreases at large values of $\Gamma/\gamma$. The bistability behavior of the coefficients ${\cal T}$, ${\cal R}$, ${\cal A}$, as a function of ${\cal J}_i/{\cal J}_0$ is never observed in agreement with results of Section \ref{bistab?}. 

\section{Summary and conclusions\label{sec:summary}}

To summarize, we have theoretically studied the nonlinear electrodynamic response of graphene at low (microwave, terahertz) frequencies, $\hbar\omega\lesssim 2E_F$, not using the perturbative theory. We have defined a set of generalized, electric field dependent conductivities $\sigma_{(2n+1)\omega,\omega}(\omega\tau,{\cal F}_\tau)$, Eq. (\ref{generConduct}), which determine the current response of graphene at the frequency harmonics $(2n+1)\omega$ if the incident electromagnetic radiation is characterized by a single harmonic $\omega$. We have investigated the conductivities $\sigma_{(2n+1)\omega,\omega}(\omega\tau,{\cal F}_\tau)$ as functions of the frequency parameter $\omega\tau$ and of the external electric field parameter ${\cal F}_\tau$, Eq. (\ref{Ftau}). We have shown that, while at low fields, at ${\cal F}_\tau\lesssim 1$, the higher harmonics amplitudes fall down very strongly with the harmonics number $n$ (proportional to ${\cal F}_\tau^n$), at higher fields, ${\cal F}_\tau\gtrsim 1$, the higher harmonics have much larger relative amplitudes $\propto 1/n$. 

We have also investigated the scattering-of-radiation problem and studied the transmission, reflection and absorption coefficients of the monochromatic radiation as a function of the frequency ($\omega\tau$), of the ratio $\Gamma/\gamma$ of the radiative decay rate $\Gamma$ to the scattering rate $\gamma$, and of the dimensionless intensity ${\cal J}_i/{\cal J}_0$, where the characteristic power density in graphene is defined in Eq. (\ref{J0}). We have found that at large values of ${\cal J}_i/{\cal J}_0$, the reflection and absorption coefficients strongly fall down, while the transmission of graphene substantially grows and tends to 80-90\% even if at low intensities the layer mainly reflected the radiation (at large values of $\Gamma/\gamma$) and the transmission consisted of only a few percent, see, e.g., Figures \ref{fig:TRA_Gg2}--\ref{fig:TRA_Gg5}. A strong transparency enhancement effect is thus the case in graphene at large values of the incident wave power. 

We have shown that the optical bistability effect, predicted within the perturbation theory in Ref. \cite{Peres14}, disappears if to solve the problem non-perturbatively. It should be noticed that in Ref. \cite{Sharif16} the authors claimed not only to predict but also to experimentally observe the bi- and multistability effect in exfoliated graphene. However, the theoretical part in this paper is also based on a perturbative (third-order) approach. As for the experimental data (Fig. 10 in Ref. \cite{Sharif16}), they only demonstrate a slight difference of the output-vs-input optical characteristics at the growing and decreasing input power, which does not actually look as a strong and sharp hysteresis that would be expected if the true bistability was the case.  

The presented theory substantially contributes to the further understanding of the nonlinear electrodynamic properties of graphene, now within the non-perturbative approach, and paves new ways to the development of nonlinear graphene-based optoelectronics.

\acknowledgments

I thank Nuno Peres and Antti-Pekka Jauho for discussions of some issues related to the paper \cite{Peres14}, as well as Nadja Savostianova for reading the manuscript and useful comments. The work has received funding from the European
Union's Horizon 2020 research and innovation programme GrapheneCore1 under Grant Agreement No. 696656.

\appendix 

\section{Details of solving the Boltzmann equation \label{Deriv}}

In order to derive Eq. (\ref{resultBE}) we substitute the second term in Eq. (\ref{res2}) into the Fourier expansion (\ref{Fourf}),
\be 
f(p_x,p_y,t)=\gamma \int_{-\infty}^\infty ds e^{isp_x}\tilde f_0(s,p_y) 
\int_{-\infty}^te^{-\gamma (t-t')-is \int_{t'}^t F(t'') dt''}d t'.
\ee
Substituting now for $\tilde f_0(s,p_y)$ the inverse Fourier transform of (\ref{Fourf0}) and changing the order of integration we obtain
\be 
f(p_x,p_y,t)=
\frac{\gamma}{2\pi}\int_{-\infty}^\infty dp_x' f_0(p_x',p_y)
\int_{-\infty}^te^{-\gamma (t-t')}d t'
\int_{-\infty}^\infty ds e^{is\left(p_x-p_x' - \int_{t'}^t F(t'') dt''\right)}
\ee
The integral over $ds$ gives the delta function, $\int_{-\infty}^\infty \dots ds = 2\pi \delta\left(p_x-p_x' - \int_{t'}^t F(t'') dt''\right)$. Then, integrating over $dp_x'$ we obtain Eq. (\ref{resultBE}):
\be 
f(p_x,p_y,t)=
\gamma
\int_{-\infty}^te^{-\gamma (t-t')}
 f_0\left(p_x - \int_{t'}^t F(t'') dt'',p_y\right)d t'.
\ee

In order to derive (\ref{resultSplitted}) from (\ref{resultBE}) we rewrite now the integral over $dt'$ as a sum of two integrals, the first one from $-\infty$ to 0 and the second from 0 to $t$, 
\be
f(p_x,p_y,t)=
\gamma
\int_{-\infty}^0e^{-\gamma (t-t')}
 f_0\left(p_x - \int_{t'}^t F(t'') dt'',p_y\right)d t'
+\gamma
\int_{0}^te^{-\gamma (t-t')}
 f_0\left(p_x - \int_{t'}^t F(t'') dt'',p_y\right)d t'.
\ee
In the first integral $t'<0<t$, and the lower limit in the integral over $dt''$ can be replaced by zero. Then the function $f_0$ does not depend on $t'$, and the integral over $dt'$ can be taken. In both integrals the function $F(t'')$ can be replaced by $\tilde F(t'')$, so that finally we obtain Eq. (\ref{resultSplitted}),
\ba 
f(p_x,p_y,t)&=&
e^{-\gamma t}
 f_0\left(p_x - \int_{0}^t \tilde F(t'') dt'',p_y\right)
+\gamma
\int_{0}^te^{-\gamma (t-t')}
 f_0\left(p_x - \int_{t'}^t \tilde F(t'') dt'',p_y\right)d t'.
\ea

In order to get Eq. (\ref{resultPeres}) from the exact solution (\ref{resultSplitted}) we introduce a new variable $\xi=\gamma (t- t')$ in the second integral (the first integral vanishes in the limit $\gamma t\to \infty$) and obtain 
\ba 
f(p_x,p_y,t)_{\gamma t\to\infty}&=&
 \int^{\gamma t}_0 e^{-\xi} f_0\left(p_x- \int_{t-\xi/\gamma}^t \tilde F(t'') dt'',p_y\right)   d \xi.
\ea
Equation (\ref{resultPeres}) follows from here in the limit $\gamma t\gg 1$.

\section{Asymptotes and Taylor expansions of the functions $B_{2n+1}(a)$ and of the conductivity $\sigma_{\omega,\omega}(\omega\tau,{\cal F}_\tau)$ \label{app:Bn}}

The hypergeometric function $_2F_1(\alpha,\beta,\gamma;z)$ which enters the definition (\ref{B2n+1}) of the functions $B_{2n+1}(a)$ is determined by the series
\be 
_2F_1(\alpha,\beta,\gamma;z)=\sum_{n=0}^\infty \frac{(\alpha)_n(\beta)_n}{(\gamma)_n}\frac{z^n}{n!}= 1+ \frac{\alpha \beta}{\gamma}\frac{z}{1}+ \frac{\alpha(\alpha+1)\beta(\beta+1)}{\gamma(\gamma+1)}\frac{z^2}{2!}
+\dots \label{2F1}
\ee
Both at $a\ll 1$ and at $a\gg 1$ the last argument ($z$) of this function is small, therefore to calculate the asymptotes of $B_{2n+1}(a)$ we can use a finite number of terms in the expansion (\ref{2F1}). This gives the following behavior of the functions $B_{2n+1}(a)$ at large and small values of the argument $a$.
At $a\gg 1$ we obtain
\ba 
B_{2n+1}(a)&\approx &
\frac 4{\pi a}\int_{0}^{\pi/2} \sin[(2n+1)x]    dx 
=\frac 4{\pi (2n+1)a},\ \ \ a\gg 1.\label{BlargeA}
\ea
If $a\ll 1$ we get
\be 
B_1(a)\approx 
1-\frac 3{2^5} a^2 -\frac 5{2^9} a^4+ O(a^6) ,\label{expanB1}
\ee
\be 
B_3(a)\approx 
\frac 1{2^5} a^2 + \frac 5{2^{10}} a^4+ O(a^6),\label{expanB3}
\ee
\be 
B_5(a)\approx 
-\frac 1{2^{10}} a^4 - \frac {35}{2^{16}} a^6+ O(a^8),\label{expanB5}
\ee
\be 
B_7(a)\approx 
\frac 5{2^{16}} a^6 + \frac {315}{2^{22}} a^8+ O(a^{10}).\label{expanB7}
\ee

In order to find the Taylor expansion of the conductivity $\sigma_{\omega,\omega}(\omega\tau,{\cal F}_\tau)=
\sigma_0 {\cal S}_{1}(\omega\tau,{\cal F}_\tau)$, defined in (\ref{sigma(2n+1)ww}) we substitute the expansion (\ref{expanB1}) in the definition of the function ${\cal S}_{1}(\omega\tau,{\cal F}_\tau)$ and obtain, after straightforward calculations:
\ba 
{\cal S}_{1}(\omega\tau,{\cal F}_\tau)&=&
\int^{\infty}_0 e^{-\xi} \frac{\sin(\omega\tau \xi/2)}{\omega\tau/2} B_{1}\left({\cal F}_\tau\frac{\sin(\omega\tau \xi/2)}{\omega\tau/2}\right)
e^{i\omega\tau \xi/2} d \xi 
\nonumber \\ &\approx&
\int^{\infty}_0 e^{-\xi} \frac{\sin(\omega\tau \xi/2)}{\omega\tau/2} \left[1-\frac 3{32}\left({\cal F}_\tau\frac{\sin(\omega\tau \xi/2)}{\omega\tau/2}\right)^2-\frac 5{512}\left({\cal F}_\tau\frac{\sin(\omega\tau \xi/2)}{\omega\tau/2}\right)^4+\dots\right]
e^{i\omega\tau \xi/2} d \xi 
\nonumber \\ &=&
\frac 1{1-i\omega\tau}-\frac 9{16}{\cal F}_\tau^2 \frac {1}{(1-2i\omega\tau)[1+(\omega\tau)^2]}
-\frac {75}{64}{\cal F}_\tau^4 \frac{ 1}{(1-3i\omega\tau)[1+(\omega\tau)^2][1+(2\omega\tau)^2]} 
+O\left({\cal F}_\tau^6\right).
\label{S1}
\ea

%


\begin{thebibliography}{55}%
\makeatletter
\providecommand \@ifxundefined [1]{%
 \@ifx{#1\undefined}
}%
\providecommand \@ifnum [1]{%
 \ifnum #1\expandafter \@firstoftwo
 \else \expandafter \@secondoftwo
 \fi
}%
\providecommand \@ifx [1]{%
 \ifx #1\expandafter \@firstoftwo
 \else \expandafter \@secondoftwo
 \fi
}%
\providecommand \natexlab [1]{#1}%
\providecommand \enquote  [1]{``#1''}%
\providecommand \bibnamefont  [1]{#1}%
\providecommand \bibfnamefont [1]{#1}%
\providecommand \citenamefont [1]{#1}%
\providecommand \href@noop [0]{\@secondoftwo}%
\providecommand \href [0]{\begingroup \@sanitize@url \@href}%
\providecommand \@href[1]{\@@startlink{#1}\@@href}%
\providecommand \@@href[1]{\endgroup#1\@@endlink}%
\providecommand \@sanitize@url [0]{\catcode `\\12\catcode `\$12\catcode
  `\&12\catcode `\#12\catcode `\^12\catcode `\_12\catcode `\%12\relax}%
\providecommand \@@startlink[1]{}%
\providecommand \@@endlink[0]{}%
\providecommand \url  [0]{\begingroup\@sanitize@url \@url }%
\providecommand \@url [1]{\endgroup\@href {#1}{\urlprefix }}%
\providecommand \urlprefix  [0]{URL }%
\providecommand \Eprint [0]{\href }%
\providecommand \doibase [0]{http://dx.doi.org/}%
\providecommand \selectlanguage [0]{\@gobble}%
\providecommand \bibinfo  [0]{\@secondoftwo}%
\providecommand \bibfield  [0]{\@secondoftwo}%
\providecommand \translation [1]{[#1]}%
\providecommand \BibitemOpen [0]{}%
\providecommand \bibitemStop [0]{}%
\providecommand \bibitemNoStop [0]{.\EOS\space}%
\providecommand \EOS [0]{\spacefactor3000\relax}%
\providecommand \BibitemShut  [1]{\csname bibitem#1\endcsname}%
\let\auto@bib@innerbib\@empty
\bibitem [{\citenamefont {Castro~Neto}\ \emph {et~al.}(2009)\citenamefont
  {Castro~Neto}, \citenamefont {Guinea}, \citenamefont {Peres}, \citenamefont
  {Novoselov},\ and\ \citenamefont {Geim}}]{Neto09}%
  \BibitemOpen
  \bibfield  {author} {\bibinfo {author} {\bibfnamefont {A.~H.}\ \bibnamefont
  {Castro~Neto}}, \bibinfo {author} {\bibfnamefont {F.}~\bibnamefont {Guinea}},
  \bibinfo {author} {\bibfnamefont {N.~M.~R.}\ \bibnamefont {Peres}}, \bibinfo
  {author} {\bibfnamefont {K.~S.}\ \bibnamefont {Novoselov}}, \ and\ \bibinfo
  {author} {\bibfnamefont {A.~K.}\ \bibnamefont {Geim}},\ }\bibfield  {title}
  {\enquote {\bibinfo {title} {The electronic properties of graphene},}\
  }\href@noop {} {\bibfield  {journal} {\bibinfo  {journal} {Rev. Mod. Phys.}\
  }\textbf {\bibinfo {volume} {81}},\ \bibinfo {pages} {109--162} (\bibinfo
  {year} {2009})}\BibitemShut {NoStop}%
\bibitem [{\citenamefont {Mikhailov}(2007)}]{Mikhailov07e}%
  \BibitemOpen
  \bibfield  {author} {\bibinfo {author} {\bibfnamefont {S.~A.}\ \bibnamefont
  {Mikhailov}},\ }\bibfield  {title} {\enquote {\bibinfo {title} {Non-linear
  electromagnetic response of graphene},}\ }\href@noop {} {\bibfield  {journal}
  {\bibinfo  {journal} {Europhys. Lett.}\ }\textbf {\bibinfo {volume} {79}},\
  \bibinfo {pages} {27002} (\bibinfo {year} {2007})}\BibitemShut {NoStop}%
\bibitem [{\citenamefont {Dragoman}\ \emph {et~al.}(2010)\citenamefont
  {Dragoman}, \citenamefont {Neculoiu}, \citenamefont {Deligeorgis},
  \citenamefont {Konstantinidis}, \citenamefont {Dragoman}, \citenamefont
  {Cismaru}, \citenamefont {Muller},\ and\ \citenamefont {Plana}}]{Dragoman10}%
  \BibitemOpen
  \bibfield  {author} {\bibinfo {author} {\bibfnamefont {M.}~\bibnamefont
  {Dragoman}}, \bibinfo {author} {\bibfnamefont {D.}~\bibnamefont {Neculoiu}},
  \bibinfo {author} {\bibfnamefont {G.}~\bibnamefont {Deligeorgis}}, \bibinfo
  {author} {\bibfnamefont {G.}~\bibnamefont {Konstantinidis}}, \bibinfo
  {author} {\bibfnamefont {D.}~\bibnamefont {Dragoman}}, \bibinfo {author}
  {\bibfnamefont {A.}~\bibnamefont {Cismaru}}, \bibinfo {author} {\bibfnamefont
  {A.~A.}\ \bibnamefont {Muller}}, \ and\ \bibinfo {author} {\bibfnamefont
  {R.}~\bibnamefont {Plana}},\ }\bibfield  {title} {\enquote {\bibinfo {title}
  {Millimeter-wave generation via frequency multiplication in graphene},}\
  }\href@noop {} {\bibfield  {journal} {\bibinfo  {journal} {Appl. Phys.
  Lett.}\ }\textbf {\bibinfo {volume} {97}},\ \bibinfo {pages} {093101}
  (\bibinfo {year} {2010})}\BibitemShut {NoStop}%
\bibitem [{\citenamefont {Bykov}\ \emph {et~al.}(2012)\citenamefont {Bykov},
  \citenamefont {Murzina}, \citenamefont {Rybin},\ and\ \citenamefont
  {Obraztsova}}]{Bykov12}%
  \BibitemOpen
  \bibfield  {author} {\bibinfo {author} {\bibfnamefont {A.~Y.}\ \bibnamefont
  {Bykov}}, \bibinfo {author} {\bibfnamefont {T.~V.}\ \bibnamefont {Murzina}},
  \bibinfo {author} {\bibfnamefont {M.~G.}\ \bibnamefont {Rybin}}, \ and\
  \bibinfo {author} {\bibfnamefont {E.~D.}\ \bibnamefont {Obraztsova}},\
  }\bibfield  {title} {\enquote {\bibinfo {title} {Second harmonic generation
  in multilayer graphene induced by direct electric current},}\ }\href@noop {}
  {\bibfield  {journal} {\bibinfo  {journal} {Phys. Rev. B}\ }\textbf {\bibinfo
  {volume} {85}},\ \bibinfo {pages} {121413(R)} (\bibinfo {year}
  {2012})}\BibitemShut {NoStop}%
\bibitem [{\citenamefont {Kumar}\ \emph {et~al.}(2013)\citenamefont {Kumar},
  \citenamefont {Kumar}, \citenamefont {Gerstenkorn}, \citenamefont {Wang},
  \citenamefont {Chiu}, \citenamefont {Smirl},\ and\ \citenamefont
  {Zhao}}]{Kumar13}%
  \BibitemOpen
  \bibfield  {author} {\bibinfo {author} {\bibfnamefont {N.}~\bibnamefont
  {Kumar}}, \bibinfo {author} {\bibfnamefont {J.}~\bibnamefont {Kumar}},
  \bibinfo {author} {\bibfnamefont {C.}~\bibnamefont {Gerstenkorn}}, \bibinfo
  {author} {\bibfnamefont {R.}~\bibnamefont {Wang}}, \bibinfo {author}
  {\bibfnamefont {H.-Y.}\ \bibnamefont {Chiu}}, \bibinfo {author}
  {\bibfnamefont {A.~L.}\ \bibnamefont {Smirl}}, \ and\ \bibinfo {author}
  {\bibfnamefont {H.}~\bibnamefont {Zhao}},\ }\bibfield  {title} {\enquote
  {\bibinfo {title} {Third harmonic generation in graphene and few-layer
  graphite films},}\ }\href@noop {} {\bibfield  {journal} {\bibinfo  {journal}
  {Phys. Rev. B}\ }\textbf {\bibinfo {volume} {87}},\ \bibinfo {pages}
  {121406(R)} (\bibinfo {year} {2013})}\BibitemShut {NoStop}%
\bibitem [{\citenamefont {Hong}\ \emph {et~al.}(2013)\citenamefont {Hong},
  \citenamefont {Dadap}, \citenamefont {Petrone}, \citenamefont {Yeh},
  \citenamefont {Hone},\ and\ \citenamefont {{Osgood, Jr.}}}]{Hong13}%
  \BibitemOpen
  \bibfield  {author} {\bibinfo {author} {\bibfnamefont {S.-Y.}\ \bibnamefont
  {Hong}}, \bibinfo {author} {\bibfnamefont {J.~I.}\ \bibnamefont {Dadap}},
  \bibinfo {author} {\bibfnamefont {N.}~\bibnamefont {Petrone}}, \bibinfo
  {author} {\bibfnamefont {P.-C.}\ \bibnamefont {Yeh}}, \bibinfo {author}
  {\bibfnamefont {J.}~\bibnamefont {Hone}}, \ and\ \bibinfo {author}
  {\bibfnamefont {R.~M.}\ \bibnamefont {{Osgood, Jr.}}},\ }\bibfield  {title}
  {\enquote {\bibinfo {title} {Optical third-harmonic generation in
  graphene},}\ }\href@noop {} {\bibfield  {journal} {\bibinfo  {journal} {Phys.
  Rev. X}\ }\textbf {\bibinfo {volume} {3}},\ \bibinfo {pages} {021014}
  (\bibinfo {year} {2013})}\BibitemShut {NoStop}%
\bibitem [{\citenamefont {An}\ \emph {et~al.}(2013)\citenamefont {An},
  \citenamefont {Nelson}, \citenamefont {Lee},\ and\ \citenamefont
  {Diebold}}]{An13}%
  \BibitemOpen
  \bibfield  {author} {\bibinfo {author} {\bibfnamefont {Y.~Q.}\ \bibnamefont
  {An}}, \bibinfo {author} {\bibfnamefont {F.}~\bibnamefont {Nelson}}, \bibinfo
  {author} {\bibfnamefont {J.~U.}\ \bibnamefont {Lee}}, \ and\ \bibinfo
  {author} {\bibfnamefont {A.~C.}\ \bibnamefont {Diebold}},\ }\bibfield
  {title} {\enquote {\bibinfo {title} {Enhanced optical second-harmonic
  generation from the current-biased graphene/{S}i{O}$_2$/{S}i(001)
  structure},}\ }\href@noop {} {\bibfield  {journal} {\bibinfo  {journal} {Nano
  Lett.}\ }\textbf {\bibinfo {volume} {13}},\ \bibinfo {pages} {2104--2109}
  (\bibinfo {year} {2013})}\BibitemShut {NoStop}%
\bibitem [{\citenamefont {An}\ \emph {et~al.}(2014)\citenamefont {An},
  \citenamefont {Rowe}, \citenamefont {Dougherty}, \citenamefont {Lee},\ and\
  \citenamefont {Diebold}}]{An14}%
  \BibitemOpen
  \bibfield  {author} {\bibinfo {author} {\bibfnamefont {Y.~Q.}\ \bibnamefont
  {An}}, \bibinfo {author} {\bibfnamefont {J.~E.}\ \bibnamefont {Rowe}},
  \bibinfo {author} {\bibfnamefont {D.~B.}\ \bibnamefont {Dougherty}}, \bibinfo
  {author} {\bibfnamefont {J.~U.}\ \bibnamefont {Lee}}, \ and\ \bibinfo
  {author} {\bibfnamefont {A.~C.}\ \bibnamefont {Diebold}},\ }\bibfield
  {title} {\enquote {\bibinfo {title} {Optical second-harmonic generation
  induced by electric current in graphene on {S}i and {S}i{C} substrates},}\
  }\href@noop {} {\bibfield  {journal} {\bibinfo  {journal} {Phys. Rev. B}\
  }\textbf {\bibinfo {volume} {89}},\ \bibinfo {pages} {115310} (\bibinfo
  {year} {2014})}\BibitemShut {NoStop}%
\bibitem [{\citenamefont {Lin}\ \emph {et~al.}(2014)\citenamefont {Lin},
  \citenamefont {Weng}, \citenamefont {Lyu}, \citenamefont {Tsai},\ and\
  \citenamefont {Su}}]{Lin14}%
  \BibitemOpen
  \bibfield  {author} {\bibinfo {author} {\bibfnamefont {K.-H.}\ \bibnamefont
  {Lin}}, \bibinfo {author} {\bibfnamefont {S.-W.}\ \bibnamefont {Weng}},
  \bibinfo {author} {\bibfnamefont {P.-W.}\ \bibnamefont {Lyu}}, \bibinfo
  {author} {\bibfnamefont {T.-R.}\ \bibnamefont {Tsai}}, \ and\ \bibinfo
  {author} {\bibfnamefont {W.-B.}\ \bibnamefont {Su}},\ }\bibfield  {title}
  {\enquote {\bibinfo {title} {Observation of optical second harmonic
  generation from suspended single-layer and bi-layer graphene},}\ }\href@noop
  {} {\bibfield  {journal} {\bibinfo  {journal} {Appl. Phys. Lett.}\ }\textbf
  {\bibinfo {volume} {105}},\ \bibinfo {pages} {151605} (\bibinfo {year}
  {2014})}\BibitemShut {NoStop}%
\bibitem [{\citenamefont {Hendry}\ \emph {et~al.}(2010)\citenamefont {Hendry},
  \citenamefont {Hale}, \citenamefont {Moger}, \citenamefont {Savchenko},\ and\
  \citenamefont {Mikhailov}}]{Hendry10}%
  \BibitemOpen
  \bibfield  {author} {\bibinfo {author} {\bibfnamefont {E.}~\bibnamefont
  {Hendry}}, \bibinfo {author} {\bibfnamefont {P.~J.}\ \bibnamefont {Hale}},
  \bibinfo {author} {\bibfnamefont {J.~J.}\ \bibnamefont {Moger}}, \bibinfo
  {author} {\bibfnamefont {A.~K.}\ \bibnamefont {Savchenko}}, \ and\ \bibinfo
  {author} {\bibfnamefont {S.~A.}\ \bibnamefont {Mikhailov}},\ }\bibfield
  {title} {\enquote {\bibinfo {title} {Coherent nonlinear optical response of
  graphene},}\ }\href@noop {} {\bibfield  {journal} {\bibinfo  {journal} {Phys.
  Rev. Lett.}\ }\textbf {\bibinfo {volume} {105}},\ \bibinfo {pages} {097401}
  (\bibinfo {year} {2010})}\BibitemShut {NoStop}%
\bibitem [{\citenamefont {Hotopan}\ \emph {et~al.}(2011)\citenamefont
  {Hotopan}, \citenamefont {{Ver Hoeye}}, \citenamefont {Vazquez},
  \citenamefont {Camblor}, \citenamefont {Fern\'andez}, \citenamefont
  {Las~Heras}, \citenamefont {\'Alvarez},\ and\ \citenamefont
  {Men\'endez}}]{Hotopan11}%
  \BibitemOpen
  \bibfield  {author} {\bibinfo {author} {\bibfnamefont {G.}~\bibnamefont
  {Hotopan}}, \bibinfo {author} {\bibfnamefont {S.}~\bibnamefont {{Ver
  Hoeye}}}, \bibinfo {author} {\bibfnamefont {C.}~\bibnamefont {Vazquez}},
  \bibinfo {author} {\bibfnamefont {R.}~\bibnamefont {Camblor}}, \bibinfo
  {author} {\bibfnamefont {M.}~\bibnamefont {Fern\'andez}}, \bibinfo {author}
  {\bibfnamefont {F.}~\bibnamefont {Las~Heras}}, \bibinfo {author}
  {\bibfnamefont {P.}~\bibnamefont {\'Alvarez}}, \ and\ \bibinfo {author}
  {\bibfnamefont {R.}~\bibnamefont {Men\'endez}},\ }\bibfield  {title}
  {\enquote {\bibinfo {title} {Millimeter wave microstrip mixer based on
  graphene},}\ }\href@noop {} {\bibfield  {journal} {\bibinfo  {journal}
  {Progress In Electromagnetic Research}\ }\textbf {\bibinfo {volume} {118}},\
  \bibinfo {pages} {57--69} (\bibinfo {year} {2011})}\BibitemShut {NoStop}%
\bibitem [{\citenamefont {Gu}\ \emph {et~al.}(2012)\citenamefont {Gu},
  \citenamefont {Petrone}, \citenamefont {McMillan}, \citenamefont {{van der
  Zande}}, \citenamefont {Yu}, \citenamefont {Lo}, \citenamefont {Kwong},
  \citenamefont {Hone},\ and\ \citenamefont {Wong}}]{Gu12}%
  \BibitemOpen
  \bibfield  {author} {\bibinfo {author} {\bibfnamefont {T.}~\bibnamefont
  {Gu}}, \bibinfo {author} {\bibfnamefont {N.}~\bibnamefont {Petrone}},
  \bibinfo {author} {\bibfnamefont {J.~F.}\ \bibnamefont {McMillan}}, \bibinfo
  {author} {\bibfnamefont {A.}~\bibnamefont {{van der Zande}}}, \bibinfo
  {author} {\bibfnamefont {M.}~\bibnamefont {Yu}}, \bibinfo {author}
  {\bibfnamefont {G.~Q.}\ \bibnamefont {Lo}}, \bibinfo {author} {\bibfnamefont
  {D.~L.}\ \bibnamefont {Kwong}}, \bibinfo {author} {\bibfnamefont
  {J.}~\bibnamefont {Hone}}, \ and\ \bibinfo {author} {\bibfnamefont {C.~W.}\
  \bibnamefont {Wong}},\ }\bibfield  {title} {\enquote {\bibinfo {title}
  {Regenerative oscillation and four-wave mixing in graphene
  optoelectronics},}\ }\href@noop {} {\bibfield  {journal} {\bibinfo  {journal}
  {Nature Photonics}\ }\textbf {\bibinfo {volume} {6}},\ \bibinfo {pages}
  {554--559} (\bibinfo {year} {2012})}\BibitemShut {NoStop}%
\bibitem [{\citenamefont {Zhang}\ \emph {et~al.}(2012)\citenamefont {Zhang},
  \citenamefont {Virally}, \citenamefont {Bao}, \citenamefont {Ping},
  \citenamefont {Massar}, \citenamefont {Godbout},\ and\ \citenamefont
  {Kockaert}}]{Zhang12}%
  \BibitemOpen
  \bibfield  {author} {\bibinfo {author} {\bibfnamefont {H.}~\bibnamefont
  {Zhang}}, \bibinfo {author} {\bibfnamefont {S.}~\bibnamefont {Virally}},
  \bibinfo {author} {\bibfnamefont {Q.~L.}\ \bibnamefont {Bao}}, \bibinfo
  {author} {\bibfnamefont {L.~K.}\ \bibnamefont {Ping}}, \bibinfo {author}
  {\bibfnamefont {S.}~\bibnamefont {Massar}}, \bibinfo {author} {\bibfnamefont
  {N.}~\bibnamefont {Godbout}}, \ and\ \bibinfo {author} {\bibfnamefont
  {P.}~\bibnamefont {Kockaert}},\ }\bibfield  {title} {\enquote {\bibinfo
  {title} {Z-scan measurement of the nonlinear refractive index of graphene},}\
  }\href@noop {} {\bibfield  {journal} {\bibinfo  {journal} {Optics Letters}\
  }\textbf {\bibinfo {volume} {37}},\ \bibinfo {pages} {1856--1858} (\bibinfo
  {year} {2012})}\BibitemShut {NoStop}%
\bibitem [{\citenamefont {Popa}\ \emph {et~al.}(2010)\citenamefont {Popa},
  \citenamefont {Sun}, \citenamefont {Torrisi}, \citenamefont {Hasan},
  \citenamefont {Wang},\ and\ \citenamefont {Ferrari}}]{Popa10}%
  \BibitemOpen
  \bibfield  {author} {\bibinfo {author} {\bibfnamefont {D.}~\bibnamefont
  {Popa}}, \bibinfo {author} {\bibfnamefont {Z.}~\bibnamefont {Sun}}, \bibinfo
  {author} {\bibfnamefont {F.}~\bibnamefont {Torrisi}}, \bibinfo {author}
  {\bibfnamefont {T.}~\bibnamefont {Hasan}}, \bibinfo {author} {\bibfnamefont
  {F.}~\bibnamefont {Wang}}, \ and\ \bibinfo {author} {\bibfnamefont {A.~C.}\
  \bibnamefont {Ferrari}},\ }\bibfield  {title} {\enquote {\bibinfo {title}
  {Sub 200 fs pulse generation from a graphene mode-locked fiber laser},}\
  }\href@noop {} {\bibfield  {journal} {\bibinfo  {journal} {Appl. Phys.
  Lett.}\ }\textbf {\bibinfo {volume} {97}},\ \bibinfo {pages} {203106}
  (\bibinfo {year} {2010})}\BibitemShut {NoStop}%
\bibitem [{\citenamefont {Popa}\ \emph {et~al.}(2011)\citenamefont {Popa},
  \citenamefont {Sun}, \citenamefont {Hasan}, \citenamefont {Torrisi},
  \citenamefont {Wang},\ and\ \citenamefont {Ferrari}}]{Popa11}%
  \BibitemOpen
  \bibfield  {author} {\bibinfo {author} {\bibfnamefont {D.}~\bibnamefont
  {Popa}}, \bibinfo {author} {\bibfnamefont {Z.}~\bibnamefont {Sun}}, \bibinfo
  {author} {\bibfnamefont {T.}~\bibnamefont {Hasan}}, \bibinfo {author}
  {\bibfnamefont {F.}~\bibnamefont {Torrisi}}, \bibinfo {author} {\bibfnamefont
  {F.}~\bibnamefont {Wang}}, \ and\ \bibinfo {author} {\bibfnamefont {A.~C.}\
  \bibnamefont {Ferrari}},\ }\bibfield  {title} {\enquote {\bibinfo {title}
  {Graphene {Q}-switched, tunable fiber laser},}\ }\href@noop {} {\bibfield
  {journal} {\bibinfo  {journal} {Appl. Phys. Lett.}\ }\textbf {\bibinfo
  {volume} {98}},\ \bibinfo {pages} {073106} (\bibinfo {year}
  {2011})}\BibitemShut {NoStop}%
\bibitem [{\citenamefont {Vermeulen}\ \emph {et~al.}(2016)\citenamefont
  {Vermeulen}, \citenamefont {Castell\'o-Lurbe}, \citenamefont {Cheng},
  \citenamefont {Pasternak}, \citenamefont {Krajewska}, \citenamefont {Ciuk},
  \citenamefont {Strupinski}, \citenamefont {Thienpont},\ and\ \citenamefont
  {Van~Erps}}]{Vermeulen16}%
  \BibitemOpen
  \bibfield  {author} {\bibinfo {author} {\bibfnamefont {N.}~\bibnamefont
  {Vermeulen}}, \bibinfo {author} {\bibfnamefont {D.}~\bibnamefont
  {Castell\'o-Lurbe}}, \bibinfo {author} {\bibfnamefont {J.~L.}\ \bibnamefont
  {Cheng}}, \bibinfo {author} {\bibfnamefont {I.}~\bibnamefont {Pasternak}},
  \bibinfo {author} {\bibfnamefont {A.}~\bibnamefont {Krajewska}}, \bibinfo
  {author} {\bibfnamefont {T.}~\bibnamefont {Ciuk}}, \bibinfo {author}
  {\bibfnamefont {W.}~\bibnamefont {Strupinski}}, \bibinfo {author}
  {\bibfnamefont {H.}~\bibnamefont {Thienpont}}, \ and\ \bibinfo {author}
  {\bibfnamefont {J.}~\bibnamefont {Van~Erps}},\ }\bibfield  {title} {\enquote
  {\bibinfo {title} {Negative {K}err nonlinearity of graphene as seen via
  chirped-pulse-pumped self-phase modulation},}\ }\href {\doibase
  10.1103/PhysRevApplied.6.044006} {\bibfield  {journal} {\bibinfo  {journal}
  {Phys. Rev. Applied}\ }\textbf {\bibinfo {volume} {6}},\ \bibinfo {pages}
  {044006} (\bibinfo {year} {2016})}\BibitemShut {NoStop}%
\bibitem [{\citenamefont {Constant}\ \emph {et~al.}(2016)\citenamefont
  {Constant}, \citenamefont {Hornett}, \citenamefont {Chang},\ and\
  \citenamefont {Hendry}}]{Constant16}%
  \BibitemOpen
  \bibfield  {author} {\bibinfo {author} {\bibfnamefont {T.~J.}\ \bibnamefont
  {Constant}}, \bibinfo {author} {\bibfnamefont {S.~M.}\ \bibnamefont
  {Hornett}}, \bibinfo {author} {\bibfnamefont {D.~E.}\ \bibnamefont {Chang}},
  \ and\ \bibinfo {author} {\bibfnamefont {E.}~\bibnamefont {Hendry}},\
  }\bibfield  {title} {\enquote {\bibinfo {title} {All-optical generation of
  surface plasmons in graphene},}\ }\href@noop {} {\bibfield  {journal}
  {\bibinfo  {journal} {Nat. Phys.}\ }\textbf {\bibinfo {volume} {12}},\
  \bibinfo {pages} {124--127} (\bibinfo {year} {2016})}\BibitemShut {NoStop}%
\bibitem [{\citenamefont {Mics}\ \emph {et~al.}(2015)\citenamefont {Mics},
  \citenamefont {Tielrooij}, \citenamefont {Parvez}, \citenamefont {Jensen},
  \citenamefont {Ivanov}, \citenamefont {Feng}, \citenamefont {M\"ullen},
  \citenamefont {Bonn},\ and\ \citenamefont {Turchinovich}}]{Mics15}%
  \BibitemOpen
  \bibfield  {author} {\bibinfo {author} {\bibfnamefont {Z.}~\bibnamefont
  {Mics}}, \bibinfo {author} {\bibfnamefont {K.-J.}\ \bibnamefont {Tielrooij}},
  \bibinfo {author} {\bibfnamefont {K.}~\bibnamefont {Parvez}}, \bibinfo
  {author} {\bibfnamefont {S.~A.}\ \bibnamefont {Jensen}}, \bibinfo {author}
  {\bibfnamefont {I.}~\bibnamefont {Ivanov}}, \bibinfo {author} {\bibfnamefont
  {X.}~\bibnamefont {Feng}}, \bibinfo {author} {\bibfnamefont {K.}~\bibnamefont
  {M\"ullen}}, \bibinfo {author} {\bibfnamefont {M.}~\bibnamefont {Bonn}}, \
  and\ \bibinfo {author} {\bibfnamefont {D.}~\bibnamefont {Turchinovich}},\
  }\bibfield  {title} {\enquote {\bibinfo {title} {Thermodynamic picture of
  ultrafast charge transport in graphene},}\ }\href@noop {} {\bibfield
  {journal} {\bibinfo  {journal} {Nature Commun.}\ }\textbf {\bibinfo {volume}
  {6}},\ \bibinfo {pages} {7655} (\bibinfo {year} {2015})}\BibitemShut
  {NoStop}%
\bibitem [{\citenamefont {Mikhailov}\ and\ \citenamefont
  {Ziegler}(2008)}]{Mikhailov08a}%
  \BibitemOpen
  \bibfield  {author} {\bibinfo {author} {\bibfnamefont {S.~A.}\ \bibnamefont
  {Mikhailov}}\ and\ \bibinfo {author} {\bibfnamefont {K.}~\bibnamefont
  {Ziegler}},\ }\bibfield  {title} {\enquote {\bibinfo {title} {Non-linear
  electromagnetic response of graphene: {F}requency multiplication and the
  self-consistent field effects},}\ }\href@noop {} {\bibfield  {journal}
  {\bibinfo  {journal} {J. Phys. Condens. Matter}\ }\textbf {\bibinfo {volume}
  {20}},\ \bibinfo {pages} {384204} (\bibinfo {year} {2008})}\BibitemShut
  {NoStop}%
\bibitem [{\citenamefont {Mikhailov}(2009{\natexlab{a}})}]{Mikhailov09a}%
  \BibitemOpen
  \bibfield  {author} {\bibinfo {author} {\bibfnamefont {S.~A.}\ \bibnamefont
  {Mikhailov}},\ }\bibfield  {title} {\enquote {\bibinfo {title} {Non-linear
  graphene optics for terahertz applications},}\ }\href@noop {} {\bibfield
  {journal} {\bibinfo  {journal} {Microelectron. J.}\ }\textbf {\bibinfo
  {volume} {40}},\ \bibinfo {pages} {712--715} (\bibinfo {year}
  {2009}{\natexlab{a}})}\BibitemShut {NoStop}%
\bibitem [{\citenamefont {Mikhailov}(2009{\natexlab{b}})}]{Mikhailov09b}%
  \BibitemOpen
  \bibfield  {author} {\bibinfo {author} {\bibfnamefont {S.~A.}\ \bibnamefont
  {Mikhailov}},\ }\bibfield  {title} {\enquote {\bibinfo {title} {Nonlinear
  cyclotron resonance of a massless quasiparticle in graphene},}\ }\href@noop
  {} {\bibfield  {journal} {\bibinfo  {journal} {Phys. Rev. B}\ }\textbf
  {\bibinfo {volume} {79}},\ \bibinfo {pages} {241309(R)} (\bibinfo {year}
  {2009}{\natexlab{b}})}\BibitemShut {NoStop}%
\bibitem [{\citenamefont {Dean}\ and\ \citenamefont {van
  Driel}(2009)}]{Dean09}%
  \BibitemOpen
  \bibfield  {author} {\bibinfo {author} {\bibfnamefont {J.~J.}\ \bibnamefont
  {Dean}}\ and\ \bibinfo {author} {\bibfnamefont {H.~M.}\ \bibnamefont {van
  Driel}},\ }\bibfield  {title} {\enquote {\bibinfo {title} {Second harmonic
  generation from graphene and graphitic film},}\ }\href@noop {} {\bibfield
  {journal} {\bibinfo  {journal} {Appl. Phys. Lett.}\ }\textbf {\bibinfo
  {volume} {95}},\ \bibinfo {pages} {261910} (\bibinfo {year}
  {2009})}\BibitemShut {NoStop}%
\bibitem [{\citenamefont {Dean}\ and\ \citenamefont {van
  Driel}(2010)}]{Dean10}%
  \BibitemOpen
  \bibfield  {author} {\bibinfo {author} {\bibfnamefont {J.~J.}\ \bibnamefont
  {Dean}}\ and\ \bibinfo {author} {\bibfnamefont {H.~M.}\ \bibnamefont {van
  Driel}},\ }\bibfield  {title} {\enquote {\bibinfo {title} {Graphene and
  few-layer graphite probed by second-harmonic generation: {T}heory and
  experiment},}\ }\href@noop {} {\bibfield  {journal} {\bibinfo  {journal}
  {Phys. Rev. B}\ }\textbf {\bibinfo {volume} {82}},\ \bibinfo {pages} {125411}
  (\bibinfo {year} {2010})}\BibitemShut {NoStop}%
\bibitem [{\citenamefont {Ishikawa}(2010)}]{Ishikawa10}%
  \BibitemOpen
  \bibfield  {author} {\bibinfo {author} {\bibfnamefont {K.~L.}\ \bibnamefont
  {Ishikawa}},\ }\bibfield  {title} {\enquote {\bibinfo {title} {Nonlinear
  optical response of graphene in time domain},}\ }\href@noop {} {\bibfield
  {journal} {\bibinfo  {journal} {Phys. Rev. B}\ }\textbf {\bibinfo {volume}
  {82}},\ \bibinfo {pages} {201402} (\bibinfo {year} {2010})}\BibitemShut
  {NoStop}%
\bibitem [{\citenamefont {Mikhailov}(2011)}]{Mikhailov11c}%
  \BibitemOpen
  \bibfield  {author} {\bibinfo {author} {\bibfnamefont {S.~A.}\ \bibnamefont
  {Mikhailov}},\ }\bibfield  {title} {\enquote {\bibinfo {title} {Theory of the
  giant plasmon-enhanced second-harmonic generation in graphene and
  semiconductor two-dimensional electron systems},}\ }\href@noop {} {\bibfield
  {journal} {\bibinfo  {journal} {Phys. Rev. B}\ }\textbf {\bibinfo {volume}
  {84}},\ \bibinfo {pages} {045432} (\bibinfo {year} {2011})}\BibitemShut
  {NoStop}%
\bibitem [{\citenamefont {Jafari}(2012)}]{Jafari12}%
  \BibitemOpen
  \bibfield  {author} {\bibinfo {author} {\bibfnamefont {S.~A.}\ \bibnamefont
  {Jafari}},\ }\bibfield  {title} {\enquote {\bibinfo {title} {Nonlinear
  optical response in gapped graphene},}\ }\href@noop {} {\bibfield  {journal}
  {\bibinfo  {journal} {J. Phys. Condens. Matter}\ }\textbf {\bibinfo {volume}
  {24}},\ \bibinfo {pages} {205802} (\bibinfo {year} {2012})}\BibitemShut
  {NoStop}%
\bibitem [{\citenamefont {Mikhailov}\ and\ \citenamefont
  {Beba}(2012)}]{Mikhailov12c}%
  \BibitemOpen
  \bibfield  {author} {\bibinfo {author} {\bibfnamefont {S.~A.}\ \bibnamefont
  {Mikhailov}}\ and\ \bibinfo {author} {\bibfnamefont {D.}~\bibnamefont
  {Beba}},\ }\bibfield  {title} {\enquote {\bibinfo {title} {Nonlinear
  broadening of the plasmon linewidth in a graphene stripe},}\ }\href@noop {}
  {\bibfield  {journal} {\bibinfo  {journal} {New J. Phys.}\ }\textbf {\bibinfo
  {volume} {14}},\ \bibinfo {pages} {115024} (\bibinfo {year}
  {2012})}\BibitemShut {NoStop}%
\bibitem [{\citenamefont {Avetissian}\ \emph {et~al.}(2013)\citenamefont
  {Avetissian}, \citenamefont {Mkrtchian}, \citenamefont {Batrakov},
  \citenamefont {Maksimenko},\ and\ \citenamefont {Hoffmann}}]{Avetissian13}%
  \BibitemOpen
  \bibfield  {author} {\bibinfo {author} {\bibfnamefont {H.~K.}\ \bibnamefont
  {Avetissian}}, \bibinfo {author} {\bibfnamefont {G.~F.}\ \bibnamefont
  {Mkrtchian}}, \bibinfo {author} {\bibfnamefont {K.~G.}\ \bibnamefont
  {Batrakov}}, \bibinfo {author} {\bibfnamefont {S.~A.}\ \bibnamefont
  {Maksimenko}}, \ and\ \bibinfo {author} {\bibfnamefont {A.}~\bibnamefont
  {Hoffmann}},\ }\bibfield  {title} {\enquote {\bibinfo {title} {Multiphoton
  resonant excitations and high-harmonic generation in bilayer graphene},}\
  }\href@noop {} {\bibfield  {journal} {\bibinfo  {journal} {Phys. Rev. B}\
  }\textbf {\bibinfo {volume} {88}},\ \bibinfo {pages} {165411} (\bibinfo
  {year} {2013})}\BibitemShut {NoStop}%
\bibitem [{\citenamefont {Mikhailov}(2013)}]{Mikhailov13c}%
  \BibitemOpen
  \bibfield  {author} {\bibinfo {author} {\bibfnamefont {S.~A.}\ \bibnamefont
  {Mikhailov}},\ }\bibfield  {title} {\enquote {\bibinfo {title}
  {Electromagnetic nonlinearities in graphene},}\ }in\ \href@noop {} {\emph
  {\bibinfo {booktitle} {Carbon nanotubes and graphene for photonic
  applications}}},\ \bibinfo {editor} {edited by\ \bibinfo {editor}
  {\bibfnamefont {Shinji}\ \bibnamefont {Yamashita}}, \bibinfo {editor}
  {\bibfnamefont {Yahachi}\ \bibnamefont {Saito}}, \ and\ \bibinfo {editor}
  {\bibfnamefont {Jong~Hyun}\ \bibnamefont {Choi}}}\ (\bibinfo  {publisher}
  {Woodhead Publishing Limited},\ \bibinfo {address} {Oxford, Cambridge,
  Philadelphia, New Delhi},\ \bibinfo {year} {2013})\ Chap.~\bibinfo {chapter}
  {7}, pp.\ \bibinfo {pages} {171--219}\BibitemShut {NoStop}%
\bibitem [{\citenamefont {Cheng}\ \emph
  {et~al.}(2014{\natexlab{a}})\citenamefont {Cheng}, \citenamefont
  {Vermeulen},\ and\ \citenamefont {Sipe}}]{Cheng14a}%
  \BibitemOpen
  \bibfield  {author} {\bibinfo {author} {\bibfnamefont {J.~L.}\ \bibnamefont
  {Cheng}}, \bibinfo {author} {\bibfnamefont {N.}~\bibnamefont {Vermeulen}}, \
  and\ \bibinfo {author} {\bibfnamefont {J.~E.}\ \bibnamefont {Sipe}},\
  }\bibfield  {title} {\enquote {\bibinfo {title} {Third order optical
  nonlinearity of graphene},}\ }\href@noop {} {\bibfield  {journal} {\bibinfo
  {journal} {New J. Phys.}\ }\textbf {\bibinfo {volume} {16}},\ \bibinfo
  {pages} {053014} (\bibinfo {year} {2014}{\natexlab{a}})}\BibitemShut
  {NoStop}%
\bibitem [{\citenamefont {Cheng}\ \emph
  {et~al.}(2014{\natexlab{b}})\citenamefont {Cheng}, \citenamefont
  {Vermeulen},\ and\ \citenamefont {Sipe}}]{Cheng14b}%
  \BibitemOpen
  \bibfield  {author} {\bibinfo {author} {\bibfnamefont {J.~L.}\ \bibnamefont
  {Cheng}}, \bibinfo {author} {\bibfnamefont {N.}~\bibnamefont {Vermeulen}}, \
  and\ \bibinfo {author} {\bibfnamefont {J.~E.}\ \bibnamefont {Sipe}},\
  }\bibfield  {title} {\enquote {\bibinfo {title} {Dc current induced second
  order optical nonlinearity in graphene},}\ }\href@noop {} {\bibfield
  {journal} {\bibinfo  {journal} {Optics Express}\ }\textbf {\bibinfo {volume}
  {22}},\ \bibinfo {pages} {15868--15876} (\bibinfo {year}
  {2014}{\natexlab{b}})}\BibitemShut {NoStop}%
\bibitem [{\citenamefont {Yao}\ \emph {et~al.}(2014)\citenamefont {Yao},
  \citenamefont {Tokman},\ and\ \citenamefont {Belyanin}}]{Yao14}%
  \BibitemOpen
  \bibfield  {author} {\bibinfo {author} {\bibfnamefont {X.}~\bibnamefont
  {Yao}}, \bibinfo {author} {\bibfnamefont {M.}~\bibnamefont {Tokman}}, \ and\
  \bibinfo {author} {\bibfnamefont {A.}~\bibnamefont {Belyanin}},\ }\bibfield
  {title} {\enquote {\bibinfo {title} {Efficient nonlinear generation of {TH}z
  plasmons in graphene and topological insulators},}\ }\href@noop {} {\bibfield
   {journal} {\bibinfo  {journal} {Phys. Rev. Lett.}\ }\textbf {\bibinfo
  {volume} {112}},\ \bibinfo {pages} {055501} (\bibinfo {year}
  {2014})}\BibitemShut {NoStop}%
\bibitem [{\citenamefont {Smirnova}\ \emph {et~al.}(2014)\citenamefont
  {Smirnova}, \citenamefont {Shadrivov}, \citenamefont {Miroshnichenko},
  \citenamefont {Smirnov},\ and\ \citenamefont {Kivshar}}]{Smirnova14}%
  \BibitemOpen
  \bibfield  {author} {\bibinfo {author} {\bibfnamefont {D.~A.}\ \bibnamefont
  {Smirnova}}, \bibinfo {author} {\bibfnamefont {I.~V.}\ \bibnamefont
  {Shadrivov}}, \bibinfo {author} {\bibfnamefont {A.~E.}\ \bibnamefont
  {Miroshnichenko}}, \bibinfo {author} {\bibfnamefont {A.~I.}\ \bibnamefont
  {Smirnov}}, \ and\ \bibinfo {author} {\bibfnamefont {Y.~S.}\ \bibnamefont
  {Kivshar}},\ }\bibfield  {title} {\enquote {\bibinfo {title} {Second-harmonic
  generation by a graphene nanoparticle},}\ }\href@noop {} {\bibfield
  {journal} {\bibinfo  {journal} {Phys. Rev. B}\ }\textbf {\bibinfo {volume}
  {90}},\ \bibinfo {pages} {035412} (\bibinfo {year} {2014})}\BibitemShut
  {NoStop}%
\bibitem [{\citenamefont {Peres}\ \emph {et~al.}(2014)\citenamefont {Peres},
  \citenamefont {Bludov}, \citenamefont {Santos}, \citenamefont {Jauho},\ and\
  \citenamefont {Vasilevskiy}}]{Peres14}%
  \BibitemOpen
  \bibfield  {author} {\bibinfo {author} {\bibfnamefont {N.~M.~R.}\
  \bibnamefont {Peres}}, \bibinfo {author} {\bibfnamefont {Y.~V.}\ \bibnamefont
  {Bludov}}, \bibinfo {author} {\bibfnamefont {J.~E.}\ \bibnamefont {Santos}},
  \bibinfo {author} {\bibfnamefont {A.-P.}\ \bibnamefont {Jauho}}, \ and\
  \bibinfo {author} {\bibfnamefont {M.~I.}\ \bibnamefont {Vasilevskiy}},\
  }\bibfield  {title} {\enquote {\bibinfo {title} {Optical bistability of
  graphene in the terahertz range},}\ }\href@noop {} {\bibfield  {journal}
  {\bibinfo  {journal} {Phys. Rev. B}\ }\textbf {\bibinfo {volume} {90}},\
  \bibinfo {pages} {125425} (\bibinfo {year} {2014})}\BibitemShut {NoStop}%
\bibitem [{\citenamefont {Cox}\ and\ \citenamefont {{de
  Abajo}}(2014)}]{CoxAbajo14}%
  \BibitemOpen
  \bibfield  {author} {\bibinfo {author} {\bibfnamefont {J.~D.}\ \bibnamefont
  {Cox}}\ and\ \bibinfo {author} {\bibfnamefont {F.~J.~G.}\ \bibnamefont {{de
  Abajo}}},\ }\bibfield  {title} {\enquote {\bibinfo {title} {Electrically
  tunable nonlinear plasmonics in graphene nanoislands},}\ }\href@noop {}
  {\bibfield  {journal} {\bibinfo  {journal} {Nat. Commun.}\ }\textbf {\bibinfo
  {volume} {5}},\ \bibinfo {pages} {5725} (\bibinfo {year} {2014})}\BibitemShut
  {NoStop}%
\bibitem [{\citenamefont {Cox}\ and\ \citenamefont {{de
  Abajo}}(2015)}]{CoxAbajo15}%
  \BibitemOpen
  \bibfield  {author} {\bibinfo {author} {\bibfnamefont {J.~D.}\ \bibnamefont
  {Cox}}\ and\ \bibinfo {author} {\bibfnamefont {F.~J.~G.}\ \bibnamefont {{de
  Abajo}}},\ }\bibfield  {title} {\enquote {\bibinfo {title} {Plasmon-enhanced
  nonlinear wave mixing in nanostructured graphene},}\ }\href@noop {}
  {\bibfield  {journal} {\bibinfo  {journal} {ACS Photonics}\ }\textbf
  {\bibinfo {volume} {2}},\ \bibinfo {pages} {306--312} (\bibinfo {year}
  {2015})}\BibitemShut {NoStop}%
\bibitem [{\citenamefont {Savostianova}\ and\ \citenamefont
  {Mikhailov}(2015)}]{Savostianova15}%
  \BibitemOpen
  \bibfield  {author} {\bibinfo {author} {\bibfnamefont {N.~A.}\ \bibnamefont
  {Savostianova}}\ and\ \bibinfo {author} {\bibfnamefont {S.~A.}\ \bibnamefont
  {Mikhailov}},\ }\bibfield  {title} {\enquote {\bibinfo {title} {Giant
  enhancement of the third harmonic in graphene integrated in a layered
  structure},}\ }\href@noop {} {\bibfield  {journal} {\bibinfo  {journal}
  {Appl. Phys. Lett.}\ }\textbf {\bibinfo {volume} {107}},\ \bibinfo {pages}
  {181104} (\bibinfo {year} {2015})}\BibitemShut {NoStop}%
\bibitem [{\citenamefont {Cheng}\ \emph {et~al.}(2015)\citenamefont {Cheng},
  \citenamefont {Vermeulen},\ and\ \citenamefont {Sipe}}]{Cheng15}%
  \BibitemOpen
  \bibfield  {author} {\bibinfo {author} {\bibfnamefont {J.~L.}\ \bibnamefont
  {Cheng}}, \bibinfo {author} {\bibfnamefont {N.}~\bibnamefont {Vermeulen}}, \
  and\ \bibinfo {author} {\bibfnamefont {J.~E.}\ \bibnamefont {Sipe}},\
  }\bibfield  {title} {\enquote {\bibinfo {title} {Third-order nonlinearity of
  graphene: Effects of phenomenological relaxation and finite temperature},}\
  }\href@noop {} {\bibfield  {journal} {\bibinfo  {journal} {Phys. Rev. B}\
  }\textbf {\bibinfo {volume} {91}},\ \bibinfo {pages} {235320} (\bibinfo
  {year} {2015})}\BibitemShut {NoStop}%
\bibitem [{\citenamefont {Cheng}\ \emph
  {et~al.}(2016{\natexlab{a}})\citenamefont {Cheng}, \citenamefont
  {Vermeulen},\ and\ \citenamefont {Sipe}}]{Cheng16}%
  \BibitemOpen
  \bibfield  {author} {\bibinfo {author} {\bibfnamefont {J.~L.}\ \bibnamefont
  {Cheng}}, \bibinfo {author} {\bibfnamefont {N.}~\bibnamefont {Vermeulen}}, \
  and\ \bibinfo {author} {\bibfnamefont {J.~E.}\ \bibnamefont {Sipe}},\
  }\bibfield  {title} {\enquote {\bibinfo {title} {Erratum: Third-order
  nonlinearity of graphene: Effects of phenomenological relaxation and finite
  temperature [{P}hys. {R}ev. {B} 91, 235320 (2015)]},}\ }\href@noop {}
  {\bibfield  {journal} {\bibinfo  {journal} {Phys. Rev. B}\ }\textbf {\bibinfo
  {volume} {93}},\ \bibinfo {pages} {039904(E)} (\bibinfo {year}
  {2016}{\natexlab{a}})}\BibitemShut {NoStop}%
\bibitem [{\citenamefont {Mikhailov}(2016)}]{Mikhailov16a}%
  \BibitemOpen
  \bibfield  {author} {\bibinfo {author} {\bibfnamefont {S.~A.}\ \bibnamefont
  {Mikhailov}},\ }\bibfield  {title} {\enquote {\bibinfo {title} {Quantum
  theory of the third-order nonlinear electrodynamic effects in graphene},}\
  }\href@noop {} {\bibfield  {journal} {\bibinfo  {journal} {Phys. Rev. B}\
  }\textbf {\bibinfo {volume} {93}},\ \bibinfo {pages} {085403} (\bibinfo
  {year} {2016})}\BibitemShut {NoStop}%
\bibitem [{\citenamefont {Mikhailov}\ \emph {et~al.}(2016)\citenamefont
  {Mikhailov}, \citenamefont {Savostianova},\ and\ \citenamefont
  {Moskalenko}}]{Mikhailov16b}%
  \BibitemOpen
  \bibfield  {author} {\bibinfo {author} {\bibfnamefont {S.~A.}\ \bibnamefont
  {Mikhailov}}, \bibinfo {author} {\bibfnamefont {N.~A.}\ \bibnamefont
  {Savostianova}}, \ and\ \bibinfo {author} {\bibfnamefont {A.~S.}\
  \bibnamefont {Moskalenko}},\ }\bibfield  {title} {\enquote {\bibinfo {title}
  {Negative dynamic conductivity of a current-driven array of graphene
  nanoribbons},}\ }\href@noop {} {\bibfield  {journal} {\bibinfo  {journal}
  {Phys. Rev. B}\ }\textbf {\bibinfo {volume} {94}},\ \bibinfo {pages} {035439}
  (\bibinfo {year} {2016})}\BibitemShut {NoStop}%
\bibitem [{\citenamefont {Rostami}\ and\ \citenamefont
  {Polini}(2016)}]{Rostami16a}%
  \BibitemOpen
  \bibfield  {author} {\bibinfo {author} {\bibfnamefont {H.}~\bibnamefont
  {Rostami}}\ and\ \bibinfo {author} {\bibfnamefont {M.}~\bibnamefont
  {Polini}},\ }\bibfield  {title} {\enquote {\bibinfo {title} {Theory of
  third-harmonic generation in graphene: A diagrammatic approach},}\
  }\href@noop {} {\bibfield  {journal} {\bibinfo  {journal} {Phys. Rev. B}\
  }\textbf {\bibinfo {volume} {93}},\ \bibinfo {pages} {161411(R)} (\bibinfo
  {year} {2016})}\BibitemShut {NoStop}%
\bibitem [{\citenamefont {Cox}\ \emph {et~al.}(2016)\citenamefont {Cox},
  \citenamefont {Silviero},\ and\ \citenamefont {{de Abajo}}}]{CoxAbajo16}%
  \BibitemOpen
  \bibfield  {author} {\bibinfo {author} {\bibfnamefont {J.~D.}\ \bibnamefont
  {Cox}}, \bibinfo {author} {\bibfnamefont {I.}~\bibnamefont {Silviero}}, \
  and\ \bibinfo {author} {\bibfnamefont {F.~J.~G.}\ \bibnamefont {{de
  Abajo}}},\ }\bibfield  {title} {\enquote {\bibinfo {title} {Quantum effects
  in the nonlinear response of graphene plasmons},}\ }\href@noop {} {\bibfield
  {journal} {\bibinfo  {journal} {ACS NANO}\ }\textbf {\bibinfo {volume}
  {10}},\ \bibinfo {pages} {1995--2003} (\bibinfo {year} {2016})}\BibitemShut
  {NoStop}%
\bibitem [{\citenamefont {Marini}\ \emph {et~al.}(2016)\citenamefont {Marini},
  \citenamefont {Cox},\ and\ \citenamefont {{de Abajo}}}]{MariniAbajo16}%
  \BibitemOpen
  \bibfield  {author} {\bibinfo {author} {\bibfnamefont {A.}~\bibnamefont
  {Marini}}, \bibinfo {author} {\bibfnamefont {J.~D.}\ \bibnamefont {Cox}}, \
  and\ \bibinfo {author} {\bibfnamefont {F.~J.~G.}\ \bibnamefont {{de
  Abajo}}},\ }\bibfield  {title} {\enquote {\bibinfo {title} {Theory of
  graphene saturable absorption},}\ }\href@noop {} {\bibfield  {journal}
  {\bibinfo  {journal} {arXiv:1605.06499}\ } (\bibinfo {year}
  {2016})}\BibitemShut {NoStop}%
\bibitem [{\citenamefont {Sharif}\ \emph
  {et~al.}(2016{\natexlab{a}})\citenamefont {Sharif}, \citenamefont {Ara},
  \citenamefont {Ghafary}, \citenamefont {Salmani},\ and\ \citenamefont
  {Mohajer}}]{Sharif16}%
  \BibitemOpen
  \bibfield  {author} {\bibinfo {author} {\bibfnamefont {M.~A.}\ \bibnamefont
  {Sharif}}, \bibinfo {author} {\bibfnamefont {M.~H.~M.}\ \bibnamefont {Ara}},
  \bibinfo {author} {\bibfnamefont {B.}~\bibnamefont {Ghafary}}, \bibinfo
  {author} {\bibfnamefont {S.}~\bibnamefont {Salmani}}, \ and\ \bibinfo
  {author} {\bibfnamefont {S.}~\bibnamefont {Mohajer}},\ }\bibfield  {title}
  {\enquote {\bibinfo {title} {Experimental observation of low threshold
  optical bistability in exfoliated graphene with low oxidation degree},}\
  }\href@noop {} {\bibfield  {journal} {\bibinfo  {journal} {Opt. Mater.}\
  }\textbf {\bibinfo {volume} {53}},\ \bibinfo {pages} {80--86} (\bibinfo
  {year} {2016}{\natexlab{a}})}\BibitemShut {NoStop}%
\bibitem [{\citenamefont {Sharif}\ \emph
  {et~al.}(2016{\natexlab{b}})\citenamefont {Sharif}, \citenamefont {Ghafary},\
  and\ \citenamefont {Ara}}]{Sharif16a}%
  \BibitemOpen
  \bibfield  {author} {\bibinfo {author} {\bibfnamefont {M.~A.}\ \bibnamefont
  {Sharif}}, \bibinfo {author} {\bibfnamefont {B.}~\bibnamefont {Ghafary}}, \
  and\ \bibinfo {author} {\bibfnamefont {M.~H.~M.}\ \bibnamefont {Ara}},\
  }\bibfield  {title} {\enquote {\bibinfo {title} {A novel graphene-based
  electro-optical modulator using modulation instability},}\ }\href@noop {}
  {\bibfield  {journal} {\bibinfo  {journal} {IEEE Photonics Technology Lett.}\
  }\textbf {\bibinfo {volume} {28}},\ \bibinfo {pages} {2897 -- 2900} (\bibinfo
  {year} {2016}{\natexlab{b}})}\BibitemShut {NoStop}%
\bibitem [{\citenamefont {Cheng}\ \emph
  {et~al.}(2016{\natexlab{b}})\citenamefont {Cheng}, \citenamefont
  {Vermeulen},\ and\ \citenamefont {Sipe}}]{Cheng16b}%
  \BibitemOpen
  \bibfield  {author} {\bibinfo {author} {\bibfnamefont {J.~L.}\ \bibnamefont
  {Cheng}}, \bibinfo {author} {\bibfnamefont {N.}~\bibnamefont {Vermeulen}}, \
  and\ \bibinfo {author} {\bibfnamefont {J.~E.}\ \bibnamefont {Sipe}},\
  }\href@noop {} {\enquote {\bibinfo {title} {Forbidden second order optical
  nonlinearity of graphene},}\ } (\bibinfo {year} {2016}{\natexlab{b}}),\
  \bibinfo {note} {arXiv:1609.06413v1}\BibitemShut {NoStop}%
\bibitem [{\citenamefont {Wang}\ \emph {et~al.}(2016)\citenamefont {Wang},
  \citenamefont {Tokman},\ and\ \citenamefont {Belyanin}}]{Wang16}%
  \BibitemOpen
  \bibfield  {author} {\bibinfo {author} {\bibfnamefont {Y.}~\bibnamefont
  {Wang}}, \bibinfo {author} {\bibfnamefont {M.}~\bibnamefont {Tokman}}, \ and\
  \bibinfo {author} {\bibfnamefont {A.}~\bibnamefont {Belyanin}},\ }\bibfield
  {title} {\enquote {\bibinfo {title} {Second-order nonlinear optical response
  of graphene},}\ }\href@noop {} {\bibfield  {journal} {\bibinfo  {journal}
  {Phys. Rev. B}\ }\textbf {\bibinfo {volume} {94}},\ \bibinfo {pages} {195442}
  (\bibinfo {year} {2016})}\BibitemShut {NoStop}%
\bibitem [{\citenamefont {Tokman}\ \emph {et~al.}(2016)\citenamefont {Tokman},
  \citenamefont {Wang}, \citenamefont {Oladyshkin}, \citenamefont {Kutayiah},\
  and\ \citenamefont {Belyanin}}]{Tokman16}%
  \BibitemOpen
  \bibfield  {author} {\bibinfo {author} {\bibfnamefont {M.}~\bibnamefont
  {Tokman}}, \bibinfo {author} {\bibfnamefont {Y.}~\bibnamefont {Wang}},
  \bibinfo {author} {\bibfnamefont {I.}~\bibnamefont {Oladyshkin}}, \bibinfo
  {author} {\bibfnamefont {A.~R.}\ \bibnamefont {Kutayiah}}, \ and\ \bibinfo
  {author} {\bibfnamefont {A.}~\bibnamefont {Belyanin}},\ }\bibfield  {title}
  {\enquote {\bibinfo {title} {Laser-driven parametric instability and
  generation of entangled photon-plasmon states in graphene},}\ }\href@noop {}
  {\bibfield  {journal} {\bibinfo  {journal} {Phys. Rev. B}\ }\textbf {\bibinfo
  {volume} {93}},\ \bibinfo {pages} {235422} (\bibinfo {year}
  {2016})}\BibitemShut {NoStop}%
\bibitem [{\citenamefont {Rostami}\ \emph {et~al.}(2016)\citenamefont
  {Rostami}, \citenamefont {Katsnelson},\ and\ \citenamefont
  {Polini}}]{Rostami16b}%
  \BibitemOpen
  \bibfield  {author} {\bibinfo {author} {\bibfnamefont {H.}~\bibnamefont
  {Rostami}}, \bibinfo {author} {\bibfnamefont {M.~I.}\ \bibnamefont
  {Katsnelson}}, \ and\ \bibinfo {author} {\bibfnamefont {M.}~\bibnamefont
  {Polini}},\ }\href@noop {} {\enquote {\bibinfo {title} {Theory of plasmonic
  effects in nonlinear optics: the case of graphene.}}\ } (\bibinfo {year}
  {2016}),\ \bibinfo {note} {arXiv:1610.04854}\BibitemShut {NoStop}%
\bibitem [{\citenamefont {Glazov}\ and\ \citenamefont
  {Ganichev}(2014)}]{Glazov14}%
  \BibitemOpen
  \bibfield  {author} {\bibinfo {author} {\bibfnamefont {M.~M.}\ \bibnamefont
  {Glazov}}\ and\ \bibinfo {author} {\bibfnamefont {S.~D.}\ \bibnamefont
  {Ganichev}},\ }\bibfield  {title} {\enquote {\bibinfo {title} {High frequency
  electric field induced nonlinear effects in graphene},}\ }\href@noop {}
  {\bibfield  {journal} {\bibinfo  {journal} {Phys. Rep.}\ }\textbf {\bibinfo
  {volume} {535}},\ \bibinfo {pages} {101--138} (\bibinfo {year}
  {2014})}\BibitemShut {NoStop}%
\bibitem [{\citenamefont {Hartmann}\ \emph {et~al.}(2014)\citenamefont
  {Hartmann}, \citenamefont {Kono},\ and\ \citenamefont
  {Portnoi}}]{Hartmann14}%
  \BibitemOpen
  \bibfield  {author} {\bibinfo {author} {\bibfnamefont {R.~R.}\ \bibnamefont
  {Hartmann}}, \bibinfo {author} {\bibfnamefont {J.}~\bibnamefont {Kono}}, \
  and\ \bibinfo {author} {\bibfnamefont {M.~E.}\ \bibnamefont {Portnoi}},\
  }\bibfield  {title} {\enquote {\bibinfo {title} {Terahertz science and
  technology of carbon nanomaterials},}\ }\href@noop {} {\bibfield  {journal}
  {\bibinfo  {journal} {Nanotechnology}\ }\textbf {\bibinfo {volume} {25}},\
  \bibinfo {pages} {322001} (\bibinfo {year} {2014})}\BibitemShut {NoStop}%
\bibitem [{\citenamefont {Romanets}\ and\ \citenamefont
  {Vasko}(2010)}]{Romanets10}%
  \BibitemOpen
  \bibfield  {author} {\bibinfo {author} {\bibfnamefont {P.~N.}\ \bibnamefont
  {Romanets}}\ and\ \bibinfo {author} {\bibfnamefont {F.~T.}\ \bibnamefont
  {Vasko}},\ }\bibfield  {title} {\enquote {\bibinfo {title} {Transient
  response of intrinsic graphene under ultrafast interband excitation},}\
  }\href {\doibase 10.1103/PhysRevB.81.085421} {\bibfield  {journal} {\bibinfo
  {journal} {Phys. Rev. B}\ }\textbf {\bibinfo {volume} {81}},\ \bibinfo
  {pages} {085421} (\bibinfo {year} {2010})}\BibitemShut {NoStop}%
\bibitem [{\citenamefont {Romanets}\ and\ \citenamefont
  {Vasko}(2011)}]{Romanets11}%
  \BibitemOpen
  \bibfield  {author} {\bibinfo {author} {\bibfnamefont {P.~N.}\ \bibnamefont
  {Romanets}}\ and\ \bibinfo {author} {\bibfnamefont {F.~T.}\ \bibnamefont
  {Vasko}},\ }\bibfield  {title} {\enquote {\bibinfo {title} {Depletion of
  carriers and negative differential conductivity in intrinsic graphene under a
  dc electric field},}\ }\href {\doibase 10.1103/PhysRevB.83.205427} {\bibfield
   {journal} {\bibinfo  {journal} {Phys. Rev. B}\ }\textbf {\bibinfo {volume}
  {83}},\ \bibinfo {pages} {205427} (\bibinfo {year} {2011})}\BibitemShut
  {NoStop}%
\bibitem [{\citenamefont {Ignatov}\ and\ \citenamefont
  {Romanov}(1976)}]{Ignatov76}%
  \BibitemOpen
  \bibfield  {author} {\bibinfo {author} {\bibfnamefont {A.~A.}\ \bibnamefont
  {Ignatov}}\ and\ \bibinfo {author} {\bibfnamefont {Yu.~A.}\ \bibnamefont
  {Romanov}},\ }\bibfield  {title} {\enquote {\bibinfo {title} {Nonlinear
  electromagnetic properties of semiconductors with a superlattice},}\
  }\href@noop {} {\bibfield  {journal} {\bibinfo  {journal} {phys. stat. sol.
  (b)}\ }\textbf {\bibinfo {volume} {78}},\ \bibinfo {pages} {327--333}
  (\bibinfo {year} {1976})}\BibitemShut {NoStop}%
\end{thebibliography}

\end{document}